\documentclass[aps,prb,twocolumn,
nobalancelastpage,amsmath,amssymb,
nofootinbib,showpacs]{revtex4-2}

\usepackage{graphics}      
\usepackage{graphicx}      
\usepackage{longtable}     
\usepackage{url}           
\usepackage{bm}            
\usepackage{braket}
\usepackage{color}

\begin{document}
\title{Glide-symmetric topological crystalline insulator phase in a nonprimitive lattice}
\author{Heejae Kim$^1$ and Shuichi Murakami$^{1,2}$}
\affiliation{$^1$Department of Physics, Tokyo Institute of Technology, 2-12-1 Ookayama, Meguro-ku, Tokyo, 152-8551, Japan \\
$^2$TIES, Tokyo Institute of Technology, 2-12-1 Ookayama, Meguro-ku, Tokyo, 152-8551, Japan}
\date{\today}

\begin{abstract}
We study the topological crystalline insulator phase protected by the glide symmetry, which is characterized by the $Z_2$ topological number. 
In the present paper, we derive a formula for the $Z_2$ topological invariant protected by glide symmetry in a nonprimitive lattice,
from that in a primitive lattice.
We establish a formula for the glide-$Z_2$ invariant for the space group No.~$\bm{{\it 9}}$ with glide symmetry in the base-centered lattice, by 
folding the Brillouin zone into that of the primitive lattice where the formula for the glide-$Z_2$ invariant is known. The formula is written 
in terms of integrals of the Berry curvatures and Berry phases in the $k$-space.
We also derive a formula of the glide-$Z_2$ invariant
when the inversion symmetry is added, and the space group becomes $\bm{{\it 15}}$. This reduces the formula into the Fu-Kane-like formula, expressed in terms of the 
irreducible representations at high-symmetry points in $k$ space.
We also construct these topological invariants by the layer-construction approach, and the results completely agree with those from the $k$-space approach.
\end{abstract}

\maketitle

\section{Introduction}


The importance of topology in condensed matter physics was recognized through the discovery of quantum Hall effect \cite{Klitzing1980prl45}.
Triggered by the proposals of topological insulators (TIs) \cite{HasanKane2010RevModPhys, QiZhang2011RevModPhys}, which emerge in the presence of time-reversal symmetry, material realization of topological crystalline materials is one of the hottest topics in modern condensed matter physics.
To realize topological crystalline materials, topological crystalline insulators (TCIs) are promising candidates because of the richness and complexity of crystallography ensuring topological properties of TCIs, such as robust surface states as long as certain crystal symmetry is preserved.
TCIs have been theoretically proposed, and their properties and candidate materials have been well studied 
\cite{Fang2012prb86, Slager2013nphys9, Fang2013prb87, Chiu2013prb88, Zhang2013prl111, Morimoto2013prb88, Jadaun2013prb88, Ueno2013prl111, Alexandradinata2014prl113, Shiozaki2014prb90, Kargarian2013prl110, Kindermann2015prl114, Weng2014prl112, Fulga2014prb89, Hsieh2014prb90, Tang2014nphys10, AndoFu2015, Chiu2016revmodphys88}.
In particular, mirror-symmetric TCI phases have attracted interest because they present the first material realization in IV-VI semiconductor SnTe, Pb$_{1-x}$Sn$_x$Te and Pb$_{1-x}$Sn$_x$Se 
\cite{Hsieh2012ncommun3, Dziawa2012nmater11, Tanaka2012nphys8, Xu2012ncommun3, Okada2013Science341, Liang2013nmat, Liu2013prb88, Sun2013prb88, Tang2014prb89, Liu2012nmater13, Ozawa2014prb90, Wrasse2014nanoletters14, Liu2015nanoletters15, Gyenis2013prb88}, which are supported by the ARPES (angle-resolved photoemission spectroscopy) experiments in the same materials.

TCI phases including the nonsymmorphic symmetry represent unique topological properties beyond topological insulating materials so far \cite{Liu2014prb90, Parameswaran2012nphys, Young2015prl115, Varjas2015prb92, Watanabe2015proc112, Po2016sciadv2, Fang2015prb91, Shiozaki2015prb91, Dong2016prb93, Chen2016prb93, Kim2016prb93, Shiozaki2016prb93, Zhao2016prb94, Chang2016nphys}.
In particuar, in magnetic systems, the Chern number and some other topological indices are 
given by integrals in $k$-space, and such topological indices are comprehensively given by the $K$-theory \cite{Shiozaki2018arXiv180206694}.
Very recently, antiferromagnetic TIs are confirmed in the MnBi$_{2n}$Te$_{3n+1}$ family of materials \cite{Zhang2019prl122, Li2019sciadv5, Vidal2019prb100, Wu2019sciadv5, Li2019prx9, Hao2019prx9, Chen2019prx9}.
Nevertheless, there are still few proposals for magnetic topological materials even though the first topological phenomenon in condensed matter physics was discovered in magnetic systems, namely the quantum Hall effect.

Distinct topological phases cannot be adiabatically deformed into each other and we can diagnose whether the system is topologically trivial or nontrivial thanks to topological invariants, such as Chern number for quantum Hall systems, $Z_2$ topological invariant for TIs, mirror Chern number for mirror-symmetric TCIs, and so on.
Namely, topological invariants which are integrals of wavefunctions encode topology of a given system, and two systems with a different value of topological invariants cannot be continuously deformed to each other.
One can classify TCI phases in crystalline materials by various methods, such as $K$-theory \cite{Shiozaki2018arXiv180206694}, topological quantum chemistry \cite{Bradlyn2017nature547}, and symmetry-based indicators \cite{Po2016sciadv2}.
Even though several expressions of important topological invariants are given in previous works, an explicit formula for each topological invariant of TCI phases does not follow immediately from these theories.
In Ref.~\onlinecite{Song2018ncommun9} the authors found expressions for symmetry-based indicators for all the space groups for systems with time-reversal symmetry, while it does not immediately tell us how the topological invariants in different space groups are mutually related.
Furthermore, expressions of topological invariants for systems without time-reversal symmetry have not been fully studied. In addition, some topological invariants such as the Chern number, the mirror Chern number and the glide-$Z_2$ invariant are expressed as an integral over the $k$ space, and how they are 
related with symmetry-based indicators should be studied separately for the individual cases. 
Thus, there is much room for further investigation for each topological invariant to apply material realization.


In this paper,
we study the $Z_2$ TCI phase protected by glide-symmetry for magnetic systems in nonprimitive lattices.
This phase is characterized by the glide-$Z_2$ topological invariant,  and its formula in primitive lattices is discussed in
Refs.~\onlinecite{Fang2015prb91, Shiozaki2015prb91,Kim2019prb100}.
Because the glide-$Z_2$ invariant is given in terms of integrals along the glide-invariant planes in momentum space,
nonprimitive lattice will alter the glide-invariant planes, and the formula of topological invariants should be altered.
We derive a formula of the glide-$Z_2$ invariant in terms of integrals in nonprimitive lattice systems, namely in the space group (SG) $\bm{{\it 9}}\ (Cc)$. 
Furthermore, we show that when inversion symmetry is added and the SG becomes $\bm{{\it 15}}\ (C2/c)$, the formula is expressed in terms of the irreducible representations at high-symmetry 
points. Together with a half of the Chern number along the glide plane, the topological invariants are $Z_2\times Z_2$, which is identical with the results of the symmetry-based indicator discussed in Refs.~\onlinecite{Po2017ncommun8, Watanabe2018sciadv4}. 
We also formulate a layer construction for these two space groups and find a perfect agreement with our theory above. 
From this results we conclude that the only topological invariants for these SGs are the glide-$Z_2$ invariant and the Chern number. 
Since the glide symmetry is contained in many space groups, this glide-symmetric TCI phase, ensured by the $Z_2$ topological invariant, can exist in many space groups as well.
Therefore, understanding of this $Z_2$ topological phase with glide symmetry will deepen our knowledge for topological phases in magnetic systems.
Our discussion in this paper is similarly applicable both to spinful and spinless systems. In some parts of the text we limit our discussion to spinless cases for 
simplicity, and extension to spinful systems is straightforward.

In this paper we study SGs $\bm{{\it 9}}\ (Cc)$ and $\bm{{\it 15}}\ (C2/c)$. To be more precise, we consider the type-I magnetic SGs (MSGs), whose symbols are the same as those for the corresponding crystallographic SGs, because they do not contain time-reversal. The SGs $\bm{{\it 9}}$ and $\bm{{\it 15}}$ refer to the type-I MSGs \#9.37 ($Cc$) and  \#15.85 ($C2/c$) in the notation of Bilbao Crystallographic Server \cite{Aroyo2006bilbao}


The present paper is organized as follows.
In Sec.~\ref{sec:glide_inversion_nonprim}, we derive the formula for the 
glide-$Z_2$ topological invariant in a nonprimitive lattice, both without
and with the inversion symmetry, corresponding to the SGs $\bm{{\it 9}}$
and $\bm{{\it 15}}$, respectively. 
This formula is also derived from the layer construction in Sec.~\ref{sec:LC}, and 
the results in Sec.~\ref{sec:glide_inversion_nonprim} and Sec.~\ref{sec:LC} completely agree with each other. 
We conclude the paper in Sec.~\ref{sec:conclusion}.

\section{$Z_2$ topological invariant for glide-symmetric magnetic systems with a nonprimitive lattice}
\label{sec:glide_inversion_nonprim}

In this section,
we focus on glide-symmetric magnetic systems in a nonprimitive lattice.
The formula of the glide-$Z_2$ invariant in the previous paper \cite{Kim2019prb100} does not apply to systems in a nonprimitive lattice, unlike those in a primitive lattice. Therefore, the formula for the 
glide-$Z_2$ invariant should be altered. 
In this section, we briefly review the glide-$Z_2$ invariant and explain our motivation.
Then, we rewrite the glide-$Z_2$ invariant in a nonprimitive lattice with glide symmetry only.

\subsection{$Z_2$ topological invariant for magnetic glide-symmetric systems}

In magnetic glide-symmetric systems, we can define the $Z_2$ topological invariant to characterize a topological crystalline insulator phase ensured by glide symmetry \cite{Fang2015prb91, Shiozaki2015prb91}.
Let us consider the glide operator
\begin{equation}
\hat{G}_y = \{ M_y | (c/2) \hat{\bm z} \} ,
\label{eq:Gy}
\end{equation}
where $M_y$ is the reflection operator with respect to the $zx$ plane, 
and $\hat{\bm z}$ is the unit vector along the $z$ axis.
Here we define $a$, $b$ and $c$ to be the lattice constants along the $x$, $y$, and $z$ axes, respectively, The SG $\bm{{\it 7}}$ ($Pc$) is generated by the glide (\ref{eq:Gy}) and translations
 $\{ E | a \hat{\bm x} \}$, $\{ E | b \hat{\bm y} \}$, and 
$\{ E | c \hat{\bm z} \}$ in a monoclinic primitive lattice, where $E$ is the identity operation. When we add another translation $\{ E | \frac{1}{2}(a \hat{\bm x}+b \hat{\bm y}) \}$, the space group 
becomes $\bm{{\it 9}}$  ($Cc$) in a monoclinic base-centered (nonprimitive) lattice. Henceforth, we set the lattice constants $a$, $b$, and $c$ to be unity for simplicity, unless otherwise specified.

\begin{figure}
\centering
\includegraphics[scale=0.31]{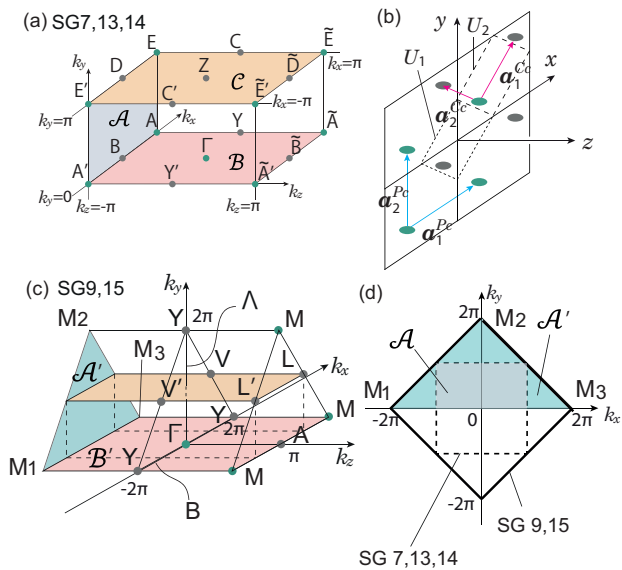}
\caption{(Color online) Unit cells and Brillouin zones for the SGs $\bm{{\it 7}}$ and $\bm{{\it 9}}$. (a) Upper half of the Brillouin zone in SGs $\bm{{\it 7}}$, $\bm{{\it 13}}$ and $\bm{{\it 14}}$. $\Gamma$, A, E, Y, Z, B, C and D denote the high-symmetry points in SG $\bm{{\it 13}}$ and SG $\bm{{\it 14}}$, while SG 
$\bm{{\it 7}}$ has no high-symmetry points, but has high-symmetry planes $k_y=0, \pi$. (b) Illustration of the relation between the primitive lattice in $\bm{{\it 7}}$ and the base-centered lattice in $\bm{{\it 9}}$ in the $xy$ plane. $U_1$ and $U_2$ denote two unit cells of the base-centered lattice $\bm{{\it 9}}$ and $U_1 + U_2$ is a unit cell of the primitive lattice $\bm{{\it 7}}$. (c) Upper half of the Brillouin zone of the monoclinic base-centered lattice in $\bm{{\it 9}}$ ($Cc$) and $\bm{{\it 15}}$ ($C2/c$).
$\Gamma, Y, V (V^\prime), A, M,$ and $L (L^\prime)$ denote the high-symmetry points in $\bm{{\it 15}}$.
The corresponding Brillouin zone of the monoclinic primitive lattice in 
SGs $\bm{{\it 7,\ 13}}$ and $\bm{{\it 14}}$ is depicted by dotted lines. 
(d) Comparison between the BZ of the primitive lattice in SGs $\bm{{\it 7}}$, $\bm{{\it 13}}$, and $\bm{{\it 14}}$ (solid line) and 
that of the base-centered lattice in SGs $\bm{{\it 9}}$, and $\bm{{\it 15}}$ (broken line).}
\label{fig:bz7-15}
\end{figure}

In SG $\bm{{\it 7}}$, we can define the glide-$Z_2$ invariant $\nu$ for such systems as \cite{Fang2015prb91, Shiozaki2015prb91,Kim2019prb100}
\begin{align}
\nu &= \frac{1}{2\pi} \left[ \int_{\mathcal{A}} F_{xy} dk_x dk_y + \int_{{\mathcal{B}}} F_{zx}^- dk_z dk_x - \int_{{\mathcal{C}}} F_{zx}^+ dk_z dk_x \right] \nonumber \\
& \quad - \frac{1}{\pi} \left( \gamma^+_{A'BA} + \gamma^-_{EDE'} \right) \pmod{2} ,
\label{eq:glide-z2}
\end{align}
where $\mathcal{A}, \mathcal{B},$ and $\mathcal{C}$ are the regions shown in Fig.~\ref{fig:bz7-15}(a).
The superscript $(\pm)$ indicates the glide sectors with eigenvalues of the glide operator $g_\pm = \pm i^f e^{-ik_z c/2}$ in spinless ($f=0$) and spinful ($f=1$) systems.
For simplicity, we limit ourselves to the spinless cases, and it is straightforward to extend our discussion to spinful cases.
$F_{ij}^\pm (\bm{k})$ is the corresponding Berry curvatures for these subspaces
\begin{equation}
F_{ij}^\pm (\bm{k}) = \partial_{k_i} A_j ^\pm (\bm{k}) - \partial_{k_j} A_i ^\pm (\bm{k}) ,
\label{eq:Fpm}
\end{equation}
and $\gamma^\pm_\lambda$ is the corresponding Berry phase within the $g_\pm$-sector along the path $\lambda$,
\begin{equation}
\gamma^\pm_\lambda (\bm{k}) = \oint_{\lambda} \bm{A}^\pm (\bm{k}) \cdot d\bm{k} ,
\end{equation}
where $\bm{A}^\pm(\bm{k})$ is the corresponding Berry connections given by the eigenstates $\ket{u^\pm_{n{\bm{k}}}}$:
\begin{equation}
\bm{A}^\pm (\bm{k}) \equiv i \sum_{n:\ \text{occ.}} \bra{u^\pm_{n{\bm k}}} \nabla_{\bm k} \ket{u^\pm_{n{\bm k}}} ,
\label{eq:Apm}
\end{equation}
where $n$ runs over occupied band indices.
The term ${{F}}_{ij}\equiv
{{F}}_{ij}^{+}+ {{F}}_{ij}^{-}$ represents the total Berry curvature for the 
occupied states.
We note that 
the Berry-phase terms $\gamma^+_{A'BA}$, $\gamma^-_{EDE'} $ in Eq.~(\ref{eq:glide-z2}) are along the paths $A'\rightarrow B\rightarrow A$, and $E\rightarrow D\rightarrow E' $, with their
the gauge taken to be periodic along the contours, meaning that
the phases of the wavefunctions at $A$ and $A'$ are the same, and likewise those at $E$ and $E'$. We also note that 
the definition of the glide-$Z_2$ invariant in Eq.~(\ref{eq:glide-z2})
is changed from that in Refs.~\onlinecite{Fang2015prb91, Shiozaki2015prb91}
for convenience in comparison with layer construction, as discussed 
in detail in Ref.~\onlinecite{Kim2019prb100}.

\subsection{Motivation}

For the glide operator in Eq.~(\ref{eq:Gy}), the Bloch Hamiltonian in momentum space satisfies
\begin{equation}
G_y \mathcal{H}(k_x, k_y, k_z) G_y^{-1} = \mathcal{H}(k_x, -k_y, k_z) ,
\label{eq:gh_comrel}
\end{equation}
where $G_y$ is the operator representing $\hat{G}_y$.
Thus, under the glide operator $G_y$, points on the $k_y = 0$ plane are invariant. Meanwhile, whether or not points on the $k_y = \pi$ plane are
invariant depends on the lattices.
In SG $\bm{{\it 7}}$, the  planes $k_y=0$ ($\mathcal{B}$) and $k_y=\pi$ ($\mathcal{C}$)
 are invariant under glide symmetry, and therefore on these planes, gilde sectors can be defined. 
On the other hand, 
in nonprimitive lattices like that in $\bm{{\it 9}}$ ($Cc$), 
$k_y = \pi$ is not glide invariant unlike SG $\bm{{\it 7}}$, and one cannot define glide sectors on this plane. Thus, it is neccessary to modify the formula of the $Z_2$ topological invariant associated with the glide-symmetry.

Let us compare $\bm{{\it 7}}$ ($Pc$) with $\bm{{\it 9}}$ ($Cc$).
Both in $\bm{{\it 7}}$ and in $\bm{{\it 9}}$, the only symmetry generator
in addition to the translations is the glide operation, Eq.~(\ref{eq:Gy}).
While $\bm{{\it 7}}$ is defined on a primitive lattice,
$\bm{{\it 9}}$ is defined on a $c$ base-centered lattice.
The primitive and reciprocal lattice vectors for $\bm{{\it 7}}$ and $\bm{{\it 9}}$ are summarized in Table~\ref{table:prim_reci_vectors}.
A half of the Brillouin zone for $\bm{{\it 7}}$ and $\bm{{\it 9}}$ is 
shown in Fig.~\ref{fig:bz7-15} (a) and (c), respectively. 
In $\bm{{\it 7}}$, every $\bm{k}$ point on the $k_y = \pi$ plane is invariant under the glide operation,
\begin{equation}
\bm{k} = g_y \bm{k} \pmod{\bm{b}_i^{Pc}} .
\label{eq:glideInv}
\end{equation}
where $\bm{b}_i^{Pc}$ $(i=1,2,3)$ are the reciprocal lattice vectors for $\bm{{\it 7}}$ ($Pc$) given in Table~\ref{table:prim_reci_vectors} and $g_y \bm{k}$ is the wavevector transformed from $\bm{k}$ by $\hat{G}_y$.
Therefore, the Bloch Hamiltonian on the $k_y = \pi$ plane can be block-diagonalized into two blocks with glide eigenvalues $g_{\pm}$,
and eigenstates on the $k_y = \pi$ plane are classified with respect to the glide sectors.
On the other hand,
$\bm{k}$ points in $\bm{{\it 9}}$ on the $k_y = \pi$ plane are not invariant under the glide operation, because $(0, 2\pi, 0)$ is not among the reciprocal vectors for $\bm{{\it 9}}$ ($Cc$).
Therefore, eigenstates on $k_y = \pi$ cannot be associated with glide sectors.
Thus, we need to alter the formula of the glide-$Z_2$ invariant for $\bm{{\it 9}}$
due to this nonprimitive nature of $\bm{{\it 9}}$.

Our strategy is the following.
To construct a formula of the glide-$Z_2$ invariant for $\bm{{\it 9}}$, we double the unit cell of $\bm{{\it 9}}$ into that of $\bm{{\it 7}}$ as shown in Fig.~\ref{fig:bz7-15}(b), leading to a folding of the BZ of $\bm{{\it 9}}$ into that of $\bm{{\it 7}}$ (see Fig.~\ref{fig:bz7-15}(d)).
Note that both the $k_y = \pi$ plane and $k_y = -\pi$ plane in $\bm{{\it 9}}$ are projected onto the same $k_y = \pi$ plane in $\bm{{\it 7}}$ by folding the BZ.
While the two points $(k_x, \pi, k_z)$ and $(k_x, -\pi, k_z)$ are distinct in $\bm{{\it 9}}$, they are mapped to the same $\bm{k}$ point in the BZ of $\bm{{\it 7}}$.
Thus, by doubling the unit cell,
we can relate the  wavefunctions between $\bm{{\it 7}}$ and $\bm{{\it 9}}$,
and we address how the terms in Eq.~(\ref{eq:glide-z2}) (for SG $\bm{{\it 7}}$) on the $k_y = \pi$ plane are described in terms of the wavefunctions in $\bm{{\it 9}}$. 
Henceforth, $\ket{u_{n\bm{k}}}$ and 
$\ket{\tilde{u}_{n\bm{k}}}$ denote a wavefunction in $\bm{{\it 9}}$ and in $\bm{{\it 7}}$, and $\mathcal{{O}}$ and $\mathcal{\tilde{O}}$ denotes an operator  in $\bm{{\it 9}}$ and in $\bm{{\it 7}}$, respectively, 
throughout the present section.

\begin{table}
\centering
\caption{Primitive lattice and reciprocal lattice vectors in monoclinic primitive and base-centered lattices. Here we set the lattice constants to be unity for simplicity.}
\begin{tabular}{ccc} 
\hline\hline
$\begin{matrix} \mathrm{space\ groups} \\ \mathrm{(lattice)} \end{matrix}$ 
  & $\bm{a}_1, \bm{a}_2, \bm{a}_3$ & $\bm{b}_1, \bm{b}_2, \bm{b}_3$ \\ \hline
$\begin{matrix}  \bm{{\it 7}}\ (Pc)\\ \mathrm{(Monoclinic} \\ \mathrm{primitive)} \end{matrix}$ & $(100), (010), (001)$ & $2\pi(100), 2\pi(010), 2\pi(001)$ \\ \hline
$\begin{matrix} \bm{{\it 9}}\ (Cc)\\ \mathrm{(Monoclinic} \\ \mathrm{base}\mbox{-}\mathrm{centered)} \end{matrix}$ &  $(\frac{1}{2}\frac{1}{2}0), (\bar{\frac{1}{2}} \frac{1}{2} 0), (001)$ & $2\pi(110), 2\pi(\bar{1}10), 2\pi(001)$ \\
\hline \hline
\end{tabular}
\label{table:prim_reci_vectors}
\end{table}

\subsection{Derivation of the formula of the $Z_2$ topological invariant for $\bm{{\it 9}}$}
\label{sec:z2-9}

We start by constructing the Hamiltonian for $\bm{{\it 7}}$ from that of $\bm{{\it 9}}$.
As depicted in Fig.~\ref{fig:bz7-15}(b),
one unit cell of $\bm{{\it 7}}$ consists of two unit cells $U_1$ and $U_2$ of $\bm{{\it 9}}$, which are related by a translation vector $\bm{a}_1 = \bm{a}_1^{Cc}
=(\frac{1}{2},\frac{1}{2},0)$.
Based on this point, we can divide the Bloch Hamiltonian for $\bm{{\it 9}}$, $\mathcal{H} (\bm{k})$, into two Hermitian parts:
\begin{equation}
\mathcal{H} (\bm{k}) = \mathcal{H}_1 (\bm{k}) + \mathcal{H}_2 (\bm{k}) 
\end{equation}
The Hamiltonian $\mathcal{H}_1 (\bm{k})$ denotes intra-unit-cell hoppings from $U_\alpha$ to $U_\alpha$ itself where $\alpha = 1,2$, while $\mathcal{H}_2 (\bm{k})$ denotes inter-unit-cell hoppings between $U_1$ and $U_2$.
Due to this division of the Hamiltonian, $\mathcal{H}_1(\bm{k})$ and $\mathcal{H}'_2(\bm{k})
\equiv\mathcal{H}_2(\bm{k})
e^{i \bm{k} \cdot \bm{a}_1}$ (not $\mathcal{H}_2 (\bm{k})$ itself) satisfy the periodicity for $\bm{{\it 7}}$ i.e., $\mathcal{H}_1 (\bm{k} + \bm{b}_i^{Pc}) = \mathcal{H}_1 (\bm{k})$ and 
$\mathcal{H}'_2 (\bm{k} + \bm{b}_i^{Pc}) = \mathcal{H}'_2 (\bm{k})$ $
 (i = 1,2,3)$ 
Then, in terms of the SG $\bm{{\it 7}}$, the corresponding Hamiltonian and the glide operator are given by
\begin{align}
\tilde{\mathcal{H}} (\bm{k}) &= 
\begin{pmatrix}
\mathcal{H}_1 (\bm{k}) & \mathcal{H}_2 (\bm{k}) e^{-i \bm{k}\cdot \bm{a}_1} \\ \mathcal{H}_2 (\bm{k}) e^{i \bm{k}\cdot \bm{a}_1} & \mathcal{H}_1 (\bm{k})
\end{pmatrix} , \\
\tilde{G}_y (\bm{k}) &= 
\begin{pmatrix}
G_y(\bm{k}) & \\ & G_y(\bm{k}) e^{-ik_y}
\end{pmatrix} ,
\end{align}
where $G_y(\bm{k})$ is the glide operator for $\bm{{\it 9}}$ satisfying ${G}_y (\bm{k}) {\mathcal{H}} (\bm{k}) {G}_y^{-1} (\bm{k}) = {\mathcal{H}} (g_y \bm{k})$. 
Therefore, we can easily see
\begin{equation}
\tilde{G}_y (\bm{k}) \tilde{\mathcal{H}} (\bm{k}) \tilde{G}_y^{-1} (\bm{k}) = \tilde{\mathcal{H}} (g_y \bm{k}).
\end{equation}
Thus we have established the Hamiltonian and the glide operator for $\bm{{\it 7}}$ out of those for $\bm{{\it 9}}$.

In order to derive a formula of the glide-$Z_2$ invariant for $\bm{{\it 9}}$, 
we need to write down wavefunctions in $\bm{{\it 7}}$ in terms of those in $\bm{{\it 9}}$.
In particular, on the $k_y = \pi$ plane,
we need wavefunctions in each of the glide sectors.
Let $\ket{u(\bm{k})}$ denote the wavefunction in $\bm{{\it 9}}$ with the energy $E_{\bm{k}}$: $\mathcal{H}(\bm{k}) \ket{u(\bm{k})} = E_{\bm{k}} \ket{u(\bm{k})}$.
Here the wavefunction $u(\bm{k})$ is periodic with respect to the Brillouin zone of the SG $\bm{{\it 9}}$. 
Then, the wavefunction for $\bm{{\it 7}}$ is given by
\begin{equation}
\ket{\tilde{u} (\bm{k})} = \frac{1}{\sqrt{2}} 
\begin{pmatrix}
1 \\ e^{i \bm{k} \cdot \bm{a}_1}
\end{pmatrix}
\ket{u (\bm{k})} ,
\end{equation}
and one can simply see that $\tilde{\mathcal{H}} (\bm{k}) \ket{\tilde{u} (\bm{k})} = E_{\bm{k}} \ket{\tilde{u} (\bm{k})}$.

Since in $\bm{{\it 7}}$, the $k_y=\pi$ plane is glide invariant, we can now construct wavefunctions in each glide sector in $\bm{{\it 7}}$ on the $k_y = \pi$ plane.
The glide symmetry guarantees
\begin{equation}
G_y(\bm{k}) \ket{u (\bm{k})} = e^{i \chi(\bm{k})} \ket{u(g_y \bm{k})} ,
\label{eq:Gu}
\end{equation}
where $\chi(\bm{k})$ is a real function.
By applying $G_y(g_y \bm{k})$, we get
\begin{equation}
e^{-ik_z} = e^{i\chi(\bm{k}) + i\chi(g_y \bm{k})},
\label{eq:chi_rel}
\end{equation}
where we used $G_y(g_y\bm{k})G_y(\bm{k})=e^{-ik_z}$ for spinless systems.
Here, we skip a discussion on spinful systems, where similar discussions can be straightforwardly developed by using the corresponding relation $G_y(g_y\bm{k})G_y(\bm{k})=-e^{-ik_z}$ instead. 
In $\bm{{\it 9}}$, $\ket{u(k_x, \pi, k_z)}$ and $G_y(k_x, \pi, k_z) \ket{u(k_x, \pi, k_z)} \propto \ket{u(k_x, -\pi, k_z)}$ reside at different wavevectors.
On the other hand, when we go to $\bm{{\it 7}}$, they correspond to the same wavevector upon halving the BZ, and their linear combinations give wavefunctions within each glide sector.
By applying $\tilde{G}_y (\bm{k})$ onto $\ket{\tilde{u}(\bm{k})}$ where $\bm{k}=\bm{k}_{\pi}\equiv (k_x, \pi, k_z)$, we have
\begin{equation}
\tilde{G}_y (\bm{k}_{\pi}) \ket{\tilde{u} (\bm{k}_{\pi})} = \frac{1}{\sqrt{2}}
\begin{pmatrix}
1 \\ -ie^{ik_x/2}
\end{pmatrix}
e^{i \chi(\bm{k}_{\pi})} \ket{u (g_y \bm{k}_{\pi})} .
\end{equation}
Thus, from the degenerate eigenstates $\ket{\tilde{u} (\bm{k}_{\pi})}$ and $\tilde{G}_y (\bm{k}) \ket{\tilde{u} (\bm{k}_{\pi})}$, we can establish eigenstates in $g_\pm$ sectors (i.e. $\tilde{G}_y  = \pm \exp (-i k_z/2)$) by linear combinations:
\begin{widetext}
\begin{align}
\ket{\tilde{u}^\pm (k_x, \pi, k_z)} &= \frac{1}{2} \left[ 
\begin{pmatrix}
1 \\ i e^{ik_x/2} 
\end{pmatrix}
\ket{u (k_x, \pi, k_z)} \pm
\begin{pmatrix}
1 \\ -i e^{ik_x/2}
\end{pmatrix} 
e^{i \chi(k_x, \pi, k_z) + ik_z/2} \ket{u (k_x, -\pi, k_z)} \right] \nonumber \\
& \quad \times \exp \left[ -i \left( \chi(\pi, \pi, k_z) + \frac{k_z}{2} \right) \frac{k_x}{2\pi} \right] \times
\begin{cases}
1 & (g_+ \ \mathrm{sector}) \\ e^{-i k_x/2} & (g_- \ \mathrm{sector})
\end{cases}
, \label{eq:evec_gpm}
\end{align}
\end{widetext}
by a straightforward calculation.
The phase factor $e^{-i (\chi(\pi, \pi, k_z) + k_z/2)\frac{k_x}{2\pi}}$ $(e^{-i (\chi(\pi, \pi, k_z) + k_z/2)\frac{k_x}{2\pi}} e^{ik_x/2} )$ is required to guarantee periodicity of the wavefunctions between $k_x = \pi$ and $k_x = -\pi$ in $\bm{{\it 7}}$, which has been imposed in the Berry-phase term in 
Eq.~(\ref{eq:glide-z2})

We plug in these eigenstates into the formula of the glide-$Z_2$ invariant of $\bm{{\it 7}}$ in Eq.~(\ref{eq:glide-z2}), with its details summarized in 
Appendix \ref{sec:Z2-9}. As a consequence, the glide-$Z_2$ invariant for $\bm{{\it 9}}$ is written as follows:
\begin{align}
\nu &= \frac{1}{2\pi} \left[ \int_{\mathcal{A}'} F_{xy} dk_xdk_y + \int_{\mathcal{B}'} \left( \frac{1}{2} F_{zx} - F_{zx}^+ \right) dk_zdk_x \right] \nonumber \\
& \quad + \frac{1}{\pi} \left( \gamma_{M_1M_2} - \gamma^+_{M_1M_3} \right) .
\label{eq:z2glide-9-2}
\end{align}
where $\mathcal{A}'=\{\bm{k}|k_z=-\pi,\ 0<k_y,\ k_x+k_y<2\pi,\ k_y-k_x<2\pi\}$ and $\mathcal{B}'=\{\bm{k}|k_y=0,\ -2\pi<k_x<2\pi,\ -\pi<k_z<\pi\}$ are depicted in Fig.~\ref{fig:bz7-15}(c) and $M_1(-2\pi,0,-\pi)$, $M_2(0,2\pi,-\pi)$, and $M_3(2\pi,0,-\pi)$,
Note that $\nu$ is gauge invariant as expected.

\subsection{Topological invariants for $\bm{{\it 15}}$}
Next, we consider what happens when the inversion symmetry is added. 
There is only one way to add  inversion symmetry $\hat{I}=\{I|\bm{0}\}$ into $\bm{{\it 9}}$, and the space group 
becomes $\bm{{\it 15}}$. In this subsection, we derive the formula of the glide-$Z_2$ invariant for $\bm{{\it 15}}$ in two ways; one is to exploit a sewing matrix introduced in Ref.~\onlinecite{Fang2012prb86},
and the other is to fold the Brillouin zone of $\bm{{\it 15}}$ into that of $\bm{{\it 13}}$ and use the formula for the glide-$Z_2$ invariant in $\bm{{\it 13}}$,
where $\bm{{\it 13}}$ consists of the glide symmetry and inversion symmetry in a 
primitive lattice.
Through these two different methods, we will show that the glide-$Z_2$ invariant $\nu$ for $\bm{{\it 15}}$ is given by
\begin{equation}
(-1)^{\nu} = \prod_{i \in \mathrm{occ}} \zeta^+_i (\Gamma) \xi_i (\mathrm{V}) \frac{\xi_i (\mathrm{Y})}{\zeta^+_i (\mathrm{Y})} ,
\end{equation}
where $\zeta_i^+$ is a $C_2$ eigenvalue in the $g_+$ sector and $\xi_i$ is an inversion parity for the $i$-th occupied state, and the product runs over the 
occupied states.
The high-symmetry points are specified in Fig.~\ref{fig:bz7-15}(c).
This formula applies both to spinless and spinful cases.
In spinless systems, this formula is rewritten as 
\begin{equation}
\nu = N_{\Gamma_2^-} (\Gamma) + N_{Y_1^-} (\mathrm{Y}) + N_{V_1^-} (\mathrm{V}) ,
\label{eq:15z2-irrep}
\end{equation}
where $N_R (P)$ is the number of occupied states at a high-symmetry point $P$ with an irrep $R$ of $\bm{{\it 15}}$ at $P$.
The irreps of $\bm{{\it 15}}$ are summarized in Appendix \ref{sec:irrep15}.

In addition, we find that the Chern number $\mathcal{C}_y$ on the $k_x$-$k_z$ plane defined by 
\begin{align}
\mathcal{C}_y &= \frac{1}{2\pi} \int_{\mathcal{B}'} F_{zx} dk_zdk_x
\end{align}
can also be related with the irreps as
\begin{equation}
\mathcal{C}_y \in 2\mathbb{Z}, \quad (-1)^{\mathcal{C}_y/2} = \prod_{i \in \mathrm{occ}} \frac{\xi_i (\mathrm{Y}) \xi_i (\mathrm{V})}{\xi_i (\mathrm{M}) \xi_i (\mathrm{L})},
\end{equation}
both for spinful and spinless systems. 
It means that $\mathcal{C}_y$  is always even, and the parity of $\mathcal{C}_y/2$ is known from the irreps of the  occupied states at high-symmetry points.

\subsubsection{Approach by using sewing matrix}
Let us start with symmetry consideration.
Adding an inversion $\hat{I} = \{ I | \bm{0} \}$ with respect to the origin to $\bm{{\it 9}}$ leads $\bm{{\it 15}}$.
Then the $C_2$ rotational symmetry around the axis $x = 0, z = 1/4$,
\begin{equation}
\hat{C}_2 = \hat{G}_y \hat{I} = \left\{ C_{2y} | \hat{\bm{z}}/2 \right\} ,
\end{equation} 
is also added to symmetry operations.
It satisfies the relation
\begin{equation}
\hat{C}_2 \hat{G}_y = \hat{G}_y \hat{C}_2 \{ E | \hat{\bm{z}} \} .
\label{eq:CR15}
\end{equation}

We now derive a formula for the glide-$Z_2$ invariant for $\bm{{\it 15}}$ from that of 
$\bm{{\it 9}}$ in Eq.~(\ref{eq:z2glide-9-2}) with the help of the additional symmetries.
We will calculate $(-1)^{\nu} = e^{i\pi\nu}$ with $\nu$ given by Eq.~(\ref{eq:z2glide-9-2}) term by term.
We find $\int_{\mathcal{A}'} F_{xy} dk_xdk_y = 0$ from $C_2$ symmetry, and
by using the sewing matrix \cite{Fang2012prb86,Kim2019prb100}, we obtain
\begin{equation}
e^{i\gamma_{M_1M_2}} = \prod_{i \in \mathrm{occ}} \frac{\xi_i (L)}{\xi_i (M)},
\end{equation}
from inversion symmetry.
Since the glide sector is unchanged upon the $C_2$ rotation on the glide-invariant plane $k_y = 0$, we have
\begin{equation}
\exp \left[ -\frac{i}{2} \int_{\mathcal{B}'} F^+_{zx} dk_z dk_x - i \gamma^+_{M_1M_3} \right] = \prod_{i \in \mathrm{occ}} \frac{\zeta_i^+ (\Gamma)}{\zeta_i^+ (Y)},
\end{equation}
which is derived by folding the region $\mathcal{B}'$ into half by the $C_2$ rotation using the sewing matrix \cite{Fang2012prb86,Kim2019prb100}.

\begin{figure}[htb]
\centering
\includegraphics[scale=0.29]{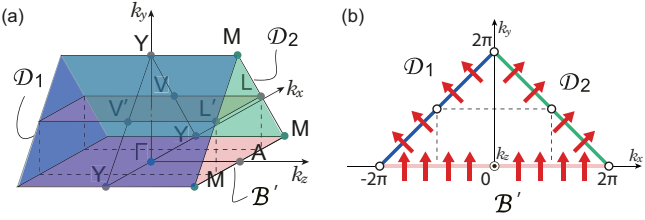}
\caption{Brillouin zone of $\bm{{\it 9}}$. (a) Upper half of the Brillouin zone of $\bm{{\it 9}}$.
(b) Schematic figure of the flux of the Berry curvatures on the  $k_z=(\mathrm{const.})$ plane denoted by red arrows.
$\mathcal{D}_1$ and $\mathcal{D}_2$ denote the rectangular regions $\mathcal{D}_1 : \{ (k-\pi, k+\pi, k_z) | -\pi < k <  \pi, -\pi < k_z < \pi \}$, $\mathcal{D}_2 : \{ (k+\pi, \pi -k, k_z) | -\pi < k <  \pi, -\pi < k_z < \pi \}$, respectively.
Integrals of the Berry curvature on $\mathcal{D}_1$ and on $\mathcal{D}_2$ are the same due to the twofold rotational symmetry $C_{2z}$. 
Because the Berry curvature is divergence free in an insulating system, its integral on $\mathcal{B}$ is equal to the sum of the those on $\mathcal{D}_1$ and on $\mathcal{D}_2$.
}
\label{fig:Berry_flux15}
\end{figure}

Finally, we study the term of the Chern number $\mathcal{C}_y=\frac{1}{2\pi}\int_{\mathcal{B}'}F_{zx}dk_zdk_x$ in Eq.~(\ref{eq:z2glide-9-2}).
The key is the fact that the Berry curvature 
is divergence free in an insulating system; in other words, the divergence of the Berry curvature can be nonzero only when the gap is closed.
Thus, from the Gauss theorem,
the Chern number $\mathcal{C}_y$ on the $k_y = 0$ plane is equal to the sum of 
the integral of the Berry curvature over the planes $\mathcal{D}_1=
\{ (k - \pi, k + \pi, k_z) | -\pi < k < \pi, -\pi < k_z < \pi \}$ and $\mathcal{D}_2=\{ (k + \pi,  \pi -k, k_z) | -\pi < k < \pi, -\pi < k_z < \pi \}$, where
the rectangular regions $\mathcal{D}_1$ and $\mathcal{D}_2$ are depicted in Fig.~\ref{fig:Berry_flux15} with their
normal vectors defined as $\bm{n}_{1}=\frac{1}{\sqrt{2}}(-1,1,0)$ and $\bm{n}_2=\frac{1}{\sqrt{2}}(1,1,0)$, respectively. 
Here we define the Chern numbers $\mathcal{C}_1$ and $\mathcal{C}_2$ over the regions  $\mathcal{D}_1$ and $\mathcal{D}_2$, respectively by
\begin{align}
&\mathcal{C}_i= \frac{1}{2\pi}\int_{\mathcal{D}_i} \ \bm{\Omega}_{\bm{k}}\cdot \bm{n}\ \ \ d^2\bm{k}\ (i=1,2),
\end{align}
where $\bm{n}$ is the normal vector of the plane, and $\bm{\Omega}_{\bm{k}}=(F_{yz},F_{zx},F_{xy})$ is the Berry curvature. 
Then, we have
\begin{align}
\mathcal{C}_y=
\frac{1}{2\pi}\int_{\mathcal{B}'} F_{zx} dk_xdk_z &= \frac{1}{2\pi}\left(\int_{\mathcal{D}_1} + \int_{\mathcal{D}_2}\right)  \bm{\Omega}_{\bm{k}}\cdot \bm{n}\ d^2\bm{k} \nonumber \\
&= \mathcal{C}_1+\mathcal{C}_2=2\mathcal{C}_1,
\label{eq:chern-15-1}
\end{align}
where we have used $\mathcal{C}_1=\mathcal{C}_2$ due to the $C_2$ rotational symmetry.
Here, $\mathcal{C}_1\ (=\mathcal{C}_2)$ is an integer, with its parity given by 
\begin{align}
& e^{i\pi\mathcal{C}_y/2} =e^{i\pi\mathcal{C}_1}=\nonumber  \\
&=\exp \left[ \frac{i}{2} \int_{\mathcal{D}_1}  \bm{\Omega}_{\bm{k}}\cdot \bm{n}\ d^2\bm{k} \right] =
\prod_{i \in \mathrm{occ}} \frac{\xi_i (Y) \xi_i (V)}{\xi_i (M) \xi_i (L)},
\label{eq:chern-15-2}
\end{align}
which is derived by folding the region $\mathcal{D}_1$ into half by the inversion using the sewing matrix \cite{Fang2012prb86,Kim2019prb100}.
Therefore,
in the present case the Chern number $\mathcal{C}_y$ on the glide plane is always an even integer.
This fact that $\mathcal{C}_y$ is an even integer also follows directly from the fact that 
the product of the inversion parities at $A$, $M$, $Y$, and $\Gamma$ on the $k_y=0$ planes being always $+1$ thanks to the compatibility relation summarized in Table \ref{table:total_irreps15}
in Appendix \ref{sec:irrep15},
by noting that this product gives $(-1)^{\mathcal{C}_y}$\cite{Hughes2011prb83,Ono2018prb98}.


As a consequence, in the presence of inversion symmetry, the glide-$Z_2$ invariant for $\bm{{\it 15}}$ is rewritten as 
\begin{align}
(-1)^{\nu} &= \prod_{i \in \mathrm{occ}} \frac{\xi_i (L)}{\xi_i (M)} \frac{\zeta_i^+ (\mathrm{\Gamma})}{\zeta_i^+ (Y)} \frac{\xi_i (Y) \xi_i (V)}{\xi_i (M) \xi_i (L)} \nonumber \\
&= \prod_{i \in \mathrm{occ}} \zeta_i^+ (\mathrm{\Gamma}) \xi_i (V) \frac{\xi_i (Y)}{\zeta_i^+ (Y)} .
\label{eq:z2-15-1}
\end{align}
So far we have obtained this equation for spinless systems, and one can similarly discuss spinful systems to get the
same result Eq.~(\ref{eq:z2-15-1}).
In particular, in spinless systems, 
we have only to count the number of states giving nontrivial contribution, whose irreps are $\Gamma_2^-$ at $\Gamma$, $V_1^-$ at $V$, and $Y_1^-$ at $Y$.
The irreps at the high-symmetry points are summarized in Table~\ref{table:total_irreps15} in Appendix \ref{sec:irrep15}. 
Therefore, we eventually have a formula of $\nu$ expressed in terms of irreps for $\bm{{\it 15}}$ as
\begin{equation}
\nu = N_{\Gamma_2^-} (\Gamma) + N_{Y_1^-} (\mathrm{Y}) + N_{V_1^-} (\mathrm{V}) ,
\end{equation}
which is the same as Eq.~(\ref{eq:15z2-irrep})

\subsubsection{Folding of the Brillouin zone from $\bm{{\it 15}}$  onto $\bm{{\it 13}}$ }

Here we derive Eq.~(\ref{eq:15z2-irrep}) for spinless cases, from the formula of the glide-$Z_2$ invariant in $\bm{{\it 13}}$ we derived in Ref.~\onlinecite{Kim2019prb100},
\begin{equation}
\nu = N_{\Gamma_2^+} (\Gamma_{13}) + N_{Y_2^+} (Y_{13}) + N_{Z_2^-} (Z_{13}) + N_{C_2^-} (C_{13}) ,
\end{equation}
in which $N_R (P_{13})$ is the number of occupied states with an irreps $R$ for $\bm{{\it 13}}$ at a high-symmetry point $P_{13}$.

Systems in $\bm{{\it 15}}$ can be regarded as those in $\bm{{\it 13}}$
by choosing the primitive lattice vectors to be $\bm{a}_1 = (1,0,0), \bm{a}_2 = (0,1,0), \bm{a}_3 = (0,0,1)$.
Thereby, the unit cell is doubled from that of $\bm{{\it 15}}$, and the BZ is folded by half, as we discussed in the previous section. To distinguish the high-symmetry $k$-points in $\bm{{\it 13}}$ from those in $\bm{{\it 15}}$, we put subscripts ``$\bm{{\it 13}}$'' and ``$\bm{{\it 15}}$'' onto the symbols of the high-symmetry points.  
On the $k_z = 0$ plane, the high-symmetry points $\Gamma_{\it 15}$ and $Y_{\it 15}$ in $\bm{{\it 15}}$ are projected onto $\Gamma_{\it 13}$ in $\bm{{\it 13}}$, and $V_{\it 15}$ is projected onto $C_{\it 13}$.
Similarly, on the $k_z = \pi$ plane, $A_{\it 15}$ and $M_{\it 15}$ are projected onto $B_{\it 13}$, and $L_{\it 15}$ is projected onto $E_{\it 13}$.
Hence, the irreps for $\Gamma_{\it 15}$ and $Y_{\it 15}$ ($A_{\it 15}$ and $M_{\it 15}$) in $\bm{{\it 15}}$ should be directly associated with those for $\Gamma_{\it 13}$ ($B_{\it 13}$) in $\bm{{\it 13}}$.
On the other hand, the case for $V_{\it 15}$ ($L_{\it 15}$) is not simple because two inequivalent points $V_{\it 15}$ and $V_{\it 15}^\prime$ ($L_{\it 15}$ and $L_{\it 15}^\prime$) in $\bm{{\it 15}}$ are mapped onto the same point in $\bm{{\it 13}}$ (Fig.~\ref{fig:bz7-15}(a)(c)).

Let us consider the state with the irrep $V_1^+$ at $V_{\it 15}$ point $\ket{u_{V_1^+} (V_{\it 15})}$: ${I}\ket{u_{V_1^+} (V_{\it 15})} = +\ket{u_{V_1^+} (V_{\it 15})}$.
By introducing a corresponding state at $V'_{15}$ given by
$\ket{\phi (V_{\it 15}^\prime)} \equiv {G}_y \ket{u_{V_1^+} (V_{\it 15})}$, we have
\begin{align}
{I}\ket{\phi (V_{\it 15}^\prime)} = {I}{G}_y\ket{u_{V_1^+} (V_{\it 15})} &= {G}_y {I}\ket{u_{V_1^+} (V_{\it 15})} \nonumber \\
&= {G}_y \ket{u_{V_1^+} (V_{\it 15})} \nonumber \\
&= \ket{\phi(V_{\it 15}^\prime)} .
\end{align}
Thus, the corresponding state at $V_{\it 15}^\prime$ is also characterized by $V_1^+$.
Note that $V_{\it 15}$ and $V_{\it 15}^\prime$ in $\bm{{\it 15}}$ are simultaneously projected onto $C_{\it 13}$ in $\bm{{\it 13}}$ in the halved BZ
and that the $C_2$ operation exchanges the states at $V_{\it 15}$ and that at $V_{\it 15}^\prime$, meaning that the character for the $C_2$ at $C_{\it 13}$ is zero. 
Therefore, the irrep $V_1^+ (V_1^-)$ at $V_{\it 15}$ in $\bm{{\it 15}}$ corresponds to the irreps $C_1^+ + C_2^+ \ (C_1^- + C_2^-)$ at $C_{\it 13}$ in $\bm{{\it 13}}$.

In the same manner, we can find a similar relation between $L_{\it 15}$ in $\bm{{\it 15}}$ and $E_{\it 13}$ in $\bm{{\it 13}}$.
We consider a state $\ket{u_{L_1^+} (L_{\it 15})}$ obeying the $L^+_1$ irrep at $L_{\it 15}$: ${I} \ket{u_{L_1^+} (L_{\it 15})} = +\ket{u_{L_1^+} (L_{\it 15})}$.
By introducing a corresponding state at $L'_{15}$ given by$\ket{\phi(L'_{\it 15})} \equiv {G}_y \ket{u_{L_1^+} (L_{\it 15})}$,
it follows that
\begin{equation}
{I} \ket{\phi (L_{\it 15}^\prime)} = -{G}_y \ket{u_{L_1^+} (L_{\it 15})} = -\ket{\phi(L_{\it 15}^\prime)} .
\end{equation}
Therefore, when the state at $L_{\it 15}$ is in the $L_1^+$ irrep, the corresponding state at $L_{\it 15}^\prime$ is characterized by $L_1^-$, and vice versa.
As a result by folding the BZ by half, the states at $L_{\it 15}$ in $\bm{{\it 15}}$ are projected onto those with the 2D irrep $E_1$ at $E_{\it 13}$ in $\bm{{\it 13}}$.

Next, in order to rewrite $ N_{Y_2^+} (Y_{13})$ and $N_{Z_2^-} (Z_{13}) $, we investigate how the irreps at $Y_{\it 13}$ and $Z_{\it 13}$ in $\bm{{\it 13}}$ correspond to those in $\bm{{\it 15}}$.
Let us start with $Y_{\it 13}$ in $\bm{{\it 13}}$. This point splits into two points $ (\pm \pi,0,0)$ on the high-symmetry plane $B_{\it 15}$ in $\bm{{\it 15}}$, which is invariant under the glide operation but not under inversion and $C_2$ rotation.
The irrep $Y_2^+$ at  $Y_{\it 13}$  corresponds to the irrep $B_2$ of $\bm{{\it 15}}$ at $B_{\it 15}$, corresponding to the $g_-$ sector, 
and we need to find out the correponding irreps at a high-symmetry point in $\bm{{\it 15}}$. Let us consider how it is compatible with irreps at  $Y_{\it 15}$
In fact, from a state $|u(\bm{k})\rangle$ following the irrep $B_2$ at $\bm{k}=(k_x,0,0)$, we can build another state  $|u'(-\bm{k})\rangle|\equiv {I}|u(\bm{k})\rangle$ at $-\bm{k}$, 
which is also in the 
same irrep $B_2$. Then the states $|\psi_{\pm}\rangle=u(\bm{k})\rangle\pm {I} |u(\bm{k})$ has even (odd) parity. Thus at $Y_{15}$, these states
$|\psi_{+}\rangle$ and $|\psi_{-}\rangle$
follow the irreps $Y_2^+$ and $Y_1^-$ at $Y_{\it 15}$, respectively. Therefore, we find
\begin{equation}
N_{Y_2^+} (Y_{\it 13}) = N_{B_2} (B_{\it 15})
= N_{Y_1^-} (Y_{\it 15}) + N_{Y_2^+} (Y_{\it 15}) .
\end{equation}
Similarly, we obtain
\begin{equation}
N_{Z_2^-} (Z_{\it 13}) =
 N_{\Lambda_2} (\Lambda_{\it 15})
= N_{Y_2^+} (Y_{\it 15}) + N_{Y_2^-} (Y_{\it 15}) .
\end{equation}
Here, We have used the fact that $Z_{\it 13}$ in $\bm{{\it 13}}$ is projected onto $(0, \pm \pi, 0)$ on the high-symmetry line $\Lambda_{\it 15}$  in $\bm{{\it 15}}$ preserving the
$C_2$ rotational symmetry but not by inversion symmetry and glide symmetry.

In summary,
we express the glide-$Z_2$ invariant in $\bm{{\it 13}}$ in terms of irreps for $\bm{{\it 15}}$ from the compatibility relations,
\begin{align}
\nu &\equiv N_{\Gamma_2^+} (\Gamma_{\it 13}) + N_{Y_2^+} (Y_{\it 13}) + N_{Z_2^-} (Z_{\it 13}) + N_{C_2^-} (C_{\it 13}) \nonumber \\
&\equiv N_{\Gamma_2^+} (\Gamma_{\it 15}) + N_{Y_2^+} (\mathrm{Y}_{\it 15}) +N_{Y_1^-} (\mathrm{Y}_{\it 15})  \nonumber\\
&\hspace{2cm}+N_{Y_2^-} (\mathrm{Y}_{\it 15})+ N_{V_1^-} (\mathrm{V}_{\it 15}) \nonumber\\
&\equiv N_{\Gamma_2^-} (\Gamma_{\it 15}) + N_{Y_1^-} (\mathrm{Y}_{\it 15}) + N_{V_1^-} (\mathrm{V}_{\it 15}) \pmod{2},
\end{align}
which is the same as Eq.~(\ref{eq:15z2-irrep}).
We have used the compatibility conditions for $\bm{{\it 15}}$
summarized in the rightmost column in Table \ref{table:total_irreps15}
in Appendix \ref{sec:irrep15}.

\subsection{Relationship with previous results}

Both  in $\bm{{\it 13}}$ and in $\bm{{\it 15}}$, 
the symmetry-based indicator is ${Z}_2 \times {Z}_2$ \cite{Po2017ncommun8, Watanabe2018sciadv4}.
By comparison with our results, we conclude that one ${Z}_2$ is given by $\nu$ in Eq.~(\ref{eq:15z2-irrep}), and the other ${Z}_2$ is in fact a half of the Chern number on the glide plane $\mathcal{C}_y/2$, which is equal to $\mathcal{C}_1(=\mathcal{C}_2)$.
Furthermore, the $E_\infty^{2,0}$ term of the $K$-theory classification \cite{Shiozaki2018arXiv180206694} is the ${Z}$. 
Since it should represent a topological invariant expressed as a $k$-space 2D integral, this $Z$ topological invariant is $\mathcal{C}_y/2(=\mathcal{C}_1=\mathcal{C}_2)$.

By the way, in inversion-symmetric systems, one can define the ${Z}_4$ symmetry-based indicator $z_4$
\cite{Po2017ncommun8, Watanabe2018sciadv4,Song2018ncommun9}, 
which is defined as a half of the sum of the parity eigenvalues $(=\pm 1)$ 
over the occupied states over the time-reversal invariant momenta. 
From our results on the relations between the glide-$Z_2$ invariant in $\bm{{\it 13}}$ and that in $\bm{{\it 15}}$, we immediately have \cite{Kim2019prb100}
\begin{equation}
\nu \equiv \frac{z_4}{2} \pmod{2} .
\label{eq:nuz4-15}
\end{equation}
Here we note that $z_4$ is always an even integer in insulating systems \cite{Hughes2011prb83,Ono2018prb98}. Therefore, since 
in the present paper we have assumed the system to be insulating, the
right hand side of Eq.~(\ref{eq:nuz4-15})
 is always an integer. Thus, the topological phase with a nontrivial glide-$Z_2$ invariant $\nu=1$ is also the higher-order topological insulator phase
with $z_4=2 \pmod{4}$. We also note that in systems with time-reversal symmetry, the formula of the $Z_4$ symmetry indicator has a simiar definition, 
just by replacing the sum of the parity eigenvalues $(=\pm 1)$ 
over the occupied states by that over the occupied Kramers pairs.

\section{Layer construction}
\label{sec:LC}

In this section, we introduce a layer construction, which enables us to construct topological invariants
in a completely different way from the $k$-space approach adopted in the previous section. We here construct 
topological invariants for $\bm{{\it 9}}$ and $\bm{{\it 15}}$ by the layer construction.

\subsection{Invariants from the layer construction}

One can introduce various real-space topological invariants by the LC, similarly to 
Refs.~\onlinecite{Song2018ncommun9, Kim2019prb100}. In this LC for a SG, we introduce a
set of planes, which are located periodically to be compatible with lattice translation symmetry. 
Here we consider each plane to be a 2D Chern insulator with a Chern number equal to $+1$.
Then for each of the SG operations, such as inversion, rotation, and glide, we can 
distinguish whether the configurations of the layers can be continuously trivialized or not, when the symmetry elements of the SG except for the operation considered are ignored.
Consequently, to characterize nontrivial configurations of planes we introduce a topological invariant for each SG operation. 

To be more specific, we introduce a layer consisting of an infinite number of planes
\begin{align}
(mnl;d) &= \left \{ \bm{r} | \bm{r} \cdot (m\bm{b}_1 + n\bm{b}_2 + l\bm{b}_3 ) = 2\pi (d+q), \ q \in \mathbb{Z}\right \} \nonumber \\
&= \left \{ \bm{r} | \frac{mx}{a} + \frac{ny}{b} + \frac{lz}{c} = d+q, \ q \in \mathbb{Z} \right \} ,
\label{eq:mnl}
\end{align}
where $\bm{b}_i$'s are reciprocal vectors corresponding to the primitive vectors 
$\bm{a}_1 = (a,0,0)$, $\bm{a}_2 = (0,b,0)$, $\bm{a}_3 = (0,0,c),$ $m$, $n$, $l$ are integers and $0 \le d < 1$.
Here, in order to facilitate understanding of the LCs for readers, we adopt the primitive vectors of the primitive lattice throughout the section. 
The integer $q$ is introduced to preserve the translation symmetry.
Each plane in the layer is decorated with a 2D Chern insulator with a Chern number $+1$.
Its orientation associated with the Chern number, i.e. the orientation of the chiral edge states, is specified by the reciprocal lattice vector $\bm{G}=m\bm{b}_1 + n\bm{b}_2 + l\bm{b}_3$, giving the 
normal vector of the plane.
Depending on SGs, we minimally introduce other layers to make the set of layers to be compatible with
the given SG. Then we evaluate whether this set of layers can be trivialized while preserving the symmetry considered is preserved.  
The details are shown in Appendix \ref{sec:app-LC}. 

In $\bm{{\it 9}}$, we can find three kinds of invariants, the Chern invariants $\delta_{\mathcal{C}_i},\ i=1,2,3$ associated with $\bm{a}_{i}^{Cc}$, a glide invariant $\delta_{\mathrm{g}_y}$ for the glide operation $G_y = \{ M_y | \hat{\bm{z}}/2 \}$, and the other glide invariant $\delta_{\mathrm{g}_n}$ for the glide operation $G_n = \{ M_y | (\hat{\bm{x}}+ \hat{\bm{y}} + \hat{\bm{z}})/2 \}$, as presented in Appendix \ref{sec:app-LC}.
The glide operation $G_n (=T_1G_y)$ is known as {\it additional symmetry elements} \cite{Hahn2002ITA,Song2018ncommun9}, where $T_1$ is a translation
by $ (\hat{\bm{x}} + \hat{\bm{y}})/2$.
Here, one can show that $\delta_{\mathcal{C}_1}=\delta_{\mathcal{C}_2}$, $\delta_{\mathcal{C}_3}=0$, $\delta_{g_n}\equiv\delta_{g_y}+\delta_{\mathcal{C}_i} \pmod{2}$,
meaning that the independent topological invariants are $\delta_{\mathcal{C}_i}$ and $\delta_{g_y}$.

In $\bm{{\it 15}}$, we can define six kinds of invariants, the Chern invariants $\delta_{\mathcal{C}_i},\ i=1,2,3$, the two glide invariants
$\delta_{\mathrm{g}_y}, \delta_{\mathrm{g}_n}$, the inversion invariant $\delta_{\text{i}}$, the $C_2$ invariant $\delta_{\text{r}}$, and the $C_2$ screw invariant $\delta_{\mathrm{s}}$,
only from geometric configuration of layers
as discussed in detail in Appendix \ref{sec:app-LC}.
The Chern invariants $\delta_{\mathcal{C}_i}$ are integers, 
while the other invariants
$\delta_{\mathrm{g}_y}, \delta_{\mathrm{g}_n}, \delta_{\text{i}}$ and $\delta_{\mathrm{s}}$ are integers defined modulo 2, and $\delta_{\text{r}}$ turns out to be always trivial:
$\delta{\text{r}}=0$
Then for every LC, one can evaluate these invariants, from which we establish relations between these
invariants:
\begin{align}
&\delta_{\text{i}}\equiv\delta_{\mathrm{g}_y} \pmod{2} , \\
&\delta_{\mathrm{s}} \equiv \delta_{\mathcal{C}_1}=\delta_{\mathcal{C}_2}\pmod{2} , \\
& \delta_{\mathrm{i}} + \delta_{\mathrm{s}} \equiv \delta_{\mathrm{g}_n} \pmod{2} , \\
& \delta_{\text{r}}\equiv 0 \pmod{2},\\
&\delta_{\mathcal{C}_3}=0
\end{align}
as shown in Appendix \ref{sec:app-LC}.

Now we can compare these invariants with the glide-$Z_2$ invariant $\nu$ and the 
Chern number $\mathcal{C}_y$. One can calculate these two topological invariants for general 
layers and compare them with the 
invariants $\delta_{\mathcal{C}_i}$, $\delta_{\mathrm{g}_y}$, $\delta_{\mathrm{g}_n}$, $\delta_{\text{i}}$, $\delta_{\text{r}}$, and $\delta_{\text{s}}$ from the LC. With 
a straightforward calculation with details presented in Appendix \ref{sec:app-LC}, we get
\begin{align}
\frac{1}{2}\mathcal{C}_y=\mathcal{C}_1=\mathcal{C}_2 &= \delta_{\mathcal{C}_1}=\delta_{\mathcal{C}_2}, \label{eq:SG15Ch} \\
\nu&\equiv\delta_{\text{i}}\equiv\delta_{\text{g}_y} \pmod{2}, \label{eq:SG15nu} \\
\frac{1}{2} \mathcal{C}_y &\equiv \delta_{\mathrm{s}} \pmod{2}, \label{eq:SG15s}\\
\nu + \frac{1}{2} \mathcal{C}_y &\equiv \delta_{\mathrm{g}_n} \pmod{2} \label{eq:SG15gn}.
\end{align}
Thus the topological invariants based on the LC completely agrees with 
the two topological invariants discussed in the previous section, the glide-$Z_2$ invariant $\nu$ and the 
Chern number $\mathcal{C}_y$. 

In particular, these results in Eqs.~(\ref{eq:SG15Ch})-(\ref{eq:SG15gn}) show that the independent topological invariants are the glide-$Z_2$ invariant $\nu$ and the 
Chern number $\mathcal{C}_y$.
From these results we can demonstrate generators of LCs for nontrivial values of topological invariants,
i.e., minimal layer configurations having nontrivial 
topological invariants. They are listed in Table
\ref{table:eLCs} in Appendix \ref{sec:app-LC}.

\subsection{Tight-binding models constructed from the layer construction for $\bm{{\it 15}}$}
Here,
we study the two elementary LCs $(lmn;d)=(001;0)$ and $(\bar{1}10;0)$ as summarized in Table~\ref{table:eLCs} in Appendix \ref{sec:app-LC} for $\bm{{\it 15}}$.
These two LCs correspond to the glide-symmetric $Z_2$ magnetic TCI and the Chern insulator phases in $\bm{{\it 15}}$, respectively,
and we construct simple tight-binding models for them in order to investigate our scenario in this subsection. 
In both models, each layer is a Chern insulator with a Chern number $+1$ for $-2<m<0$, $-1$ for $0<m<2$ and zero otherwise, 
and this model shows phase transitions at $m = 0, \pm 2$. 
The details of the models are presented in Appendix~\ref{app:layer15}.

\subsubsection{Glide-symmetric $Z_2$ magnetic TCI phase: $(001;0)$}

Let us begin with the glide-symmetric $Z_2$ magnetic TCI phase.
The glide-symmetric $Z_2$ magnetic TCI phase can be realized as an alternate stacking of two 2D Chern insulator layers with opposite signs of the Chern number $\pm 1$.
Suppose layers of a 2D Chern insulator with $\mathcal{C}_z = \pm 1$ along the $xy$ plane are placed at $z = z_0 + n$ in which $n$ is an integer.
The glide symmetry then requires presence of other layers of a 2D Chern insulator with $\mathcal{C}_z = -1$ at $z = z_0 + n + 1/2$, because the glide operation changes the sign of the Chern number. This system preserves the symmetry in the SG $\bm{{\it 9}}$.
This model gives a nontrivial glide-$Z_2$ topological invariant $\nu$ for glide-symmetric systems by a direct calculation.

\begin{figure}[tb!]
\centering
\includegraphics[scale=0.3]{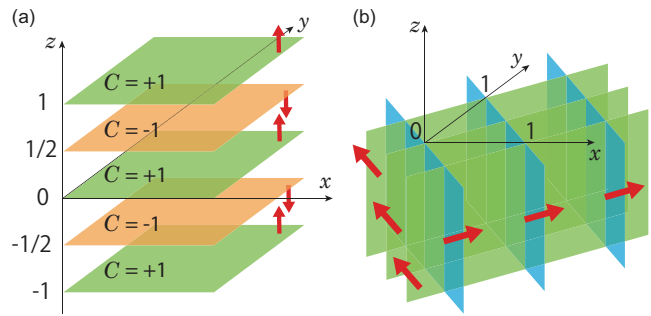}
\caption{Layer constructions for (a) $(001;0)$, showing a glide-$Z_2$ topological phase (${\nu}=1$, $\mathcal{C}_y=0$) and (b) $(\bar{1}10;0)$, showing a Chern insulator (${\nu}=0$, $\mathcal{C}_y=2$) with $\bm{{\it 15}}$. The red arrows denote the directions of the normal vector of the plane giving a positive Chern number $+1$.}
\label{fig:lc15}
\end{figure}

We now consider the case when inversion symmetry is added, leading to the SG $\bm{{\it 15}}$.
Inversion symmetry is preserved when we fix the value of $z_0$ to be $z_0 = 0$, and the configuration allowed by symmetry is the one with the 2D Chern insulators on $z = n$ with $\mathcal{C}_z = +1$ and on $z = n + 1/2$ with $\mathcal{C}_z = -1$, where $n$ is an integer (Fig.~\ref{fig:lc15}(a)).
Therefore, a representative Hamiltonian for the glide-symmetric $Z_2$ magnetic TCI phase with additional inversion symmetry is obtained as
\begin{align}
H^{\mathrm{LC}}_{\nu} &= \left( m + 2 \cos \frac{k_x}{2} \cos \frac{k_y}{2} \right) \sigma_z \nonumber\\
&+ \sin \frac{k_x}{2} \cos \frac{k_y}{2} (\sigma_x - \sigma_y) \nonumber \\
&+ \cos \frac{k_x}{2} \sin \frac{k_y}{2} (\sigma_x + \sigma_y) \tau_z,
\label{eq:15z2_layer_construction}
\end{align}
where $\sigma_j$ are the Pauli matrices corresponding to an internal degree of freedom within each layer, introduced for the purpose of 
realizing a Chern insulator, and 
$\tau_j$  are the Pauli matrices acting onto the layer index referring to 
the layers at $z=n$ and $z=n+\frac{1}{2}$.
The corresponding $k$-dependent glide operator and $k$-dependent inversion operator are
\begin{align}
G_y (k_z) &= e^{-ik_z/2} \left( \cos \frac{k_z}{2} \tau_x + \sin \frac{k_z}{2} \tau_y \right) , \label{eq:glide_operator15} \\ 
I(k_z) &= \sigma_z \begin{pmatrix} 1 & 0 \\ 0 & e^{-ik_z} \end{pmatrix} _{\tau} . \label{eq:inversion_operator15}
\end{align}
The Hamiltonian in Eq.~(\ref{eq:15z2_layer_construction}) satisfies
\begin{align}
G_y(k_z) H^{\mathrm{LC}}_{\nu} (k_x, k_y, k_z) G_y(k_z)^{-1} &= H^{\mathrm{LC}}_{\nu} (k_x, -k_y, k_z) , \label{eq:relation15_ham_glide} \\ 
I(k_z) H^{\mathrm{LC}}_{\nu} (\bm{k}) I(k_z)^{-1} &= H^{\mathrm{LC}}_{\nu} (- \bm{k}) , \label{eq:relation15_ham_inversion}
\end{align}
and these operators satisfy the algebra
\begin{equation}
G_y(-k_z) I(k_z) = e^{ik_z} I(k_z) G_y(k_z) .
\end{equation}

The Hamiltonian $H^{\mathrm{LC}}_{\nu}$ is gapped unless $m = \pm2, 0$.
The gap closes at $\Gamma$ when $m = -2$, at $V$ (and $V^\prime$) when $m = 0$, and at $Y$ when $m = 2$.
The effective Hamiltonian at the high-symmetry points $P=\Gamma$, $Y$, and $V$
on the $k_z=0$ plane in Eq.~(\ref{eq:z2-15-1}) can be written as
\begin{equation}
H^{\mathrm{LC}}_{\nu} (P) = \left( m + 2 \cos \frac{k_x}{2} \cos \frac{k_y}{2} \right) \sigma_z .
\end{equation}
The irreps of the occupied states are summarized in Table~\ref{table:LCirreps15}.
By using Eq.~(\ref{eq:15z2-irrep}), we can show that the $Z_2$ topological invariant is nontrivial (trivial) if $0 < |m| < 2$ ($|m| > 2$). In particular, when $2<m<0$, the model corresponds to the layer construction shown in Fig.~\ref{fig:lc15} as expected.

\begin{table}[tb!]
\centering
\caption{Irreducible representations for the two tight-binding models for the topological phases in $\bm{{\it 15}}$ from the layer construction.}
$$
\begin{array}{cccccc}
\hline\hline
 & \Gamma & Y & V & M & L \\ \hline
H^{\mathrm{LC}}_{\nu} (m<-2) & \Gamma_1^+  \Gamma_2^+ & Y_1^+  Y_2^+ & 2 V_1^+ & M_1 & L_1^+ L_1^- \\
H^{\mathrm{LC}}_{\nu} (-2<m<0) & \Gamma_1^-  \Gamma_2^- & Y_1^+ Y_2^+ & 2 V_1^+ & M_1 & L_1^+  L_1^- \\
H^{\mathrm{LC}}_{\nu} (0<m<2) & \Gamma_1^-  \Gamma_2^- & Y_1^+  Y_2^+ & 2 V_1^- & M_1 & L_1^+  L_1^- \\
H^{\mathrm{LC}}_{\nu} (2<m) & \Gamma_1^-  \Gamma_2^- & Y_1^- Y_2^- & 2 V_1^- & M_1 & L_1^+  L_1^- \\
\hline
H^{\mathrm{LC}}_{\mathcal{C}} (m<-2) & \Gamma_1^+  \Gamma_2^+ & Y_1^+  Y_2^+ & 2 V_1^+ & M_1 & L_1^+  L_1^- \\
H^{\mathrm{LC}}_{\mathcal{C}} (-2<m<0) & \Gamma_1^-  \Gamma_2^- & Y_1^+  Y_2^+ & V_1^+ V_1^- & M_1 & L_1^+  L_1^- \\
H^{\mathrm{LC}}_{\mathcal{C}} (0<m<2) & \Gamma_1^-  \Gamma_2^- & Y_1^-  Y_2^- & 2 V_1^- & M_1 & 2L_1^+ \\
H^{\mathrm{LC}}_{\mathcal{C}} (2<m) & \Gamma_1^-  \Gamma_2^- & Y_1^-  Y_2^- & 2 V_1^-  & M_1 & L_1^+  L_1^- \\
\hline\hline
\end{array}
$$
\label{table:LCirreps15}
\end{table}

The $z_4$ indicator is equal to {\it two} when $0<|m|<2$ and zero otherwise from Table~\ref{table:LCirreps15}.
Therefore, when $0<|m|<2$, it is a higher-order TI with the symmetry-based indicator for $\bm{{\it 2}}$ given by $(\mathbb{Z}_2, \mathbb{Z}_2, \mathbb{Z}_2, \mathbb{Z}_4) = (0,0,0;2)$  in accordance with Eq.~(\ref{eq:nuz4-15}).

\subsubsection{Chern insulator phase: ($\bar{1}10;0$)}

Next, we construct a model with a nontrivial Chern number.
In this case, we simply stack a 2D Chern insulator with a Chern number equal to $+1$ along $(\bar{1}10)$ and $(110)$ directions (Fig.~\ref{fig:lc15}(b)).
Then, we have a representative Hamiltonian
\begin{equation}
H^{\mathrm{LC}}_{\mathcal{C}} (\bm{k}) = 
\begin{pmatrix}
H^{(+)} (\bm{k}) & \\ & H^{(-)} (\bm{k}) 
\end{pmatrix}_{\tau} ,
\end{equation}
where
\begin{align}
H^{(\pm)} (\bm{k}) &= \left( m + \cos k_z + \cos \frac{k_x \pm k_y}{2} \right) \sigma_z \nonumber \\
& + \sin k_z \sigma_x + \sin \frac{k_x \pm k_y}{2} \sigma_y .
\label{eq:15chern_layer_construction}
\end{align}
The Hamiltonian $H^{\mathrm{LC}}_{\mathcal{C}}$ indeed satisfies the relations in Eqs.~(\ref{eq:relation15_ham_glide}) and (\ref{eq:relation15_ham_inversion})
under the corresponding operators given by Eqs.~(\ref{eq:glide_operator15}) and (\ref{eq:inversion_operator15}).

The irreps of the occupied states are summarized in Table~\ref{table:LCirreps15}.
By using Eq.~(\ref{eq:15z2-irrep}), we can show that the $Z_2$ topological invariant is nontrivial (trivial) if $0 < |m| < 2$ ($|m| > 2$). 
Because each layer has a Chern number $+1$ for $-2<m<0$, $-1$ for $0<m<2$ and zero otherwise,
it is a trivial insulator when $|m|>2$ and a Chern insulator with $\mathcal{C}_y = +2$ at $-2<m<0$, and  $\mathcal{C}_y = -2$ at $0<m<2$.
Meanwhile, the $Z_2$ topological invariant (\ref{eq:z2-15-1}) is always trivial, as we have expected.
In particular, when $2<m<0$, the model corresponds to the layer construction shown in Fig.~\ref{fig:lc15} as expected.

\section{Conclusion}
\label{sec:conclusion}
In the present paper, we find the formula for the topological invariant protected by glide symmetry in a nonprimitive lattice.
We establish a formula for the glide-$Z_2$ invariant for the space group $\bm{{\it 9}}$ with glide symmetry in the base-centered lattice, by 
folding its Brillouin zone into that of the primitive lattice where the formula for the glide-$Z_2$ invariant is known. The formula is written 
in terms of the integrals of the Berry curvatures and Berry phases in the $k$-space.
We then calculate the formula of the glide-$Z_2$ invariant
when the inversion symmetry is added, and the space group becomes $\bm{{\it 15}}$. This reduces the formula into the Fu-Kane-like formula, expressed in terms of the 
irreducible representations at high-symmetry points in $k$ space. This result gives an interpretation to the previous 
results on topological indices, such as symmetry-based indicators and K-theory. In particular, the symmetry-based indicator for the SG $\bm{{\it 15}}$ is
${Z}_2\times{Z}_2$, and the results in our paper show that one ${Z}_2$ is the glide-$Z_2$ topological invariant and the other is 
the parity of the half of the Chern number along the glide invariant plane $k_y=0$.

We also construct toplogical invariants by the layer-construction approach, and the results completely agree with those from the $k$-space approach in this paper.
Based on this we find two elementary layer constructions for $\bm{{\it 15}}$ in a form of tight-binding models.

In the previous paper \cite{Kim2019prb100} and in the present paper, we
discussed the fate of the glide-$Z_2$ invariant when the inversion symmetry is added. There are only three possible minimal space groups consisting of inversion symmetry and glide symmetry: the space groups $\bm{{\it 13}}$, $\bm{{\it 14}}$ and $\bm{{\it 15}}$. The former two ($\bm{{\it 13}}$ and $\bm{{\it 14}}$) are on a primitive lattice studied in the previous paper \cite{Kim2019prb100}, and the last one ($\bm{{\it 15}}$) is on a nonprimitive lattice studied in the present paper. The former two cases are distinguished by whether 
the inversion center is on a glide-mirror plane ($\bm{{\it 13}}$) or not ($\bm{{\it 14}}$).  We found that in the absence of inversion symmetry ($\bm{{\it 7}}$ and $\bm{{\it 9}}$), the glide-$Z_2$ invariant is expressed in terms of $k$-space integrals, while by adding inversion symmetry, the glide-$Z_2$ invariant is related with the irreps at high-symmetry points in $k$-space, as shown in 
the present paper and Ref.~\onlinecite{Kim2019prb100}. 
One might think that further analysis is needed for many other space groups with inversion and glide symmetries, but it is not the case. Since 
we have established that the glide-$Z_2$ invariant is related to the irreps at high-symmetry points in $k$-space in the space groups $\bm{{\it 13}}$, $\bm{{\it 14}}$ and $\bm{{\it 15}}$, it is now straightforward to apply these formulae to other space groups with inversion and glide symmetries. Thus, the topological invariants obtained in this paper should be useful for characterizing and designing 
the glide-$Z_2$ phase in systems with various kinds of particles.


\begin{acknowledgments}
This work was supported by Japan Society for the Promotion
of Science (JSPS) KAKENHI Grant No.~JP18H03678 and JP20H04633.
H. K. is supported by Japan Society for the Promotion of Science (JSPS) KAKENHI Grant-in-Aid for JSPS Fellows Grant No.~JP17J10672.
\end{acknowledgments}

\appendix

\section{Details of the derivation of the glide-$Z_2$ invariant in 
SG $\bm{{\it 9}}$}
\label{sec:Z2-9}
In this Appendix, we derive the formula of the glide-$Z_2$ invariant for the SG $\bm{{\it 9}}$ from that for $\bm{{\it 7}}$ given in
Eq.~(\ref{eq:glide-z2}).
We plug in the eigenstates of $\bm{{\it 7}}$ in Eq.~(\ref{eq:evec_gpm}) constructed from those in $\bm{{\it 9}}$ into the formula of the glide $Z_2$ invariant of $\bm{{\it 7}}$ in Eq.~(\ref{eq:glide-z2}).
In particular, we can recast the third integral in Eq.~(\ref{eq:glide-z2}) into 
\begin{align}
&\int_{\mathcal{C}}=
\int_{-\pi}^{\pi} dk_z \int_{-\pi}^{\pi} dk_x {\tilde{F}}_{zx}^+ (k_x, \pi, k_z) \nonumber \\
&= \frac{1}{2} \left( \int_{-\pi}^{\pi} dk_z \int_{-\pi}^{\pi} dk_x {F}_{zx} (k_x, \pi, k_z) \right.\nonumber\\
&+\left. \int_{-\pi}^{\pi} dk_z \int_{-\pi}^{\pi} dk_x {F}_{zx} (k_x, -\pi, k_z) \right) \nonumber \\
& \quad + \frac{1}{4} \int_{-\pi}^\pi dk_x \sum_{i\in\text{occ.}} \left( i \bra{u_i(\bm{q})} {G}_y(\bm{q}) \ket{u_i(\bm{q})} + \mathrm{c.c.} \right) ,
\label{eq:FC9-1}
\end{align}
where $\bm{q}=(k_x, \pi, -\pi)$.
Here we define $\tilde{F}_{ij}^{\pm}(\bm{k})$ and
$\tilde{\bm{A}}^{\pm}(\bm{k})$ to be the Berry cuvatures and the Berry connections within the glide $g_{\pm}$ subspaces defined in terms of the 
wavefunctions $\tilde{u}$ in SG $\bm{7}$, 
similar to Eqs.~(\ref{eq:Fpm}) and (\ref{eq:Apm}).
The Berry phase term along $\text{E}\rightarrow \text{D}\rightarrow \text{E}'$ is expressed as
\begin{align}
&-2\gamma_{EDE'}^{-}=2\oint_{-\pi}^{\pi} dk_x  \tilde{A}_x^- (k_x, \pi, -\pi) \nonumber \\
&= \oint_{-\pi}^{\pi} dk_x  {A}_x (k_x, \pi, -\pi) + \oint_{-\pi}^{\pi} dk_x {A}_x (k_x, -\pi, -\pi) \nonumber\\
&
+ \sum_{i\in\text{occ.}}(\chi_i(\pi, \pi, -\pi) + \chi_i(-\pi, \pi, -\pi))\nonumber \\
& \quad + \frac{1}{4} \int_{-\pi}^\pi dk_x \sum_{i\in\text{occ.}} \left( i \bra{u_i(\bm{q})} {G}_y(\bm{q}) \ket{u_i(\bm{q})} + \mathrm{c.c.} \right) .
\label{eq:gamma-b-9}
\end{align}

\begin{figure}[tb!]
\centering
\includegraphics[scale=0.37]{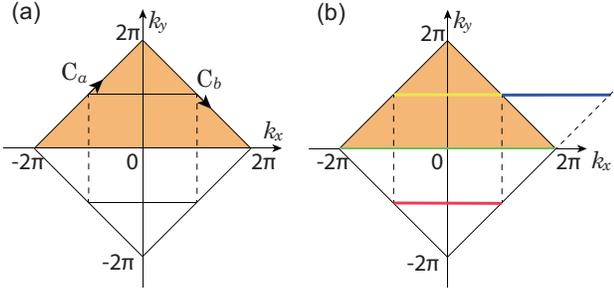}
\caption{Domain of integration in the derivation of the formula of the glide-$Z_2$ invariant in $\bm{{\it 9}}$. (a) The paths for the Berry phase terms in Eq.~(\ref{eq:ACgamma}) on 
$k_z=-\pi$. (b) Domains of integration of the
Berry curvature shown on a $k_z=(\text{const.})$ plane. In Eq.~(\ref{eq:z2glide-9-1}) the Berry curvature term is over the two regions shown as red and yellow lines. Because of the Brillouin zone periodicity, the integral over the blue region is
the same as that over the red one, and by adding the term from the yellow region, it equals to the integral over the green region
along $k_y=0$.
}
\label{fig:integral_region_9}
\end{figure}

Next, we consider the $k_z=-\pi$ plane. The term on this plane for $\bm{{\it 9}}$ is rewritten as
\begin{align}
&\int_{\mathcal{A}}=\int_{0}^{\pi} dk_y \int_{-\pi}^{\pi} dk_x {\tilde{F}}_{xy} (k_x, k_y, -\pi) \nonumber\\
&= \int_{0}^{2\pi} dk_y \int_{k_y-2\pi}^{-k_y+2\pi} dk_x {F}_{xy} (k_x, k_y, -\pi) \nonumber \\
&  +\left(\oint_{\text{C}_a}+\oint_{\text{C}_b}\right)d\bm{k}\cdot \bm{A} (\bm{k}) \nonumber \\
& - \oint_{-\pi}^{\pi} dk_x {A}_x (k_x, \pi, -\pi) - \oint_{-\pi}^{\pi} dk_x {A}_x (k_x, -\pi, -\pi),
\label{eq:FA9-1}
\end{align}
where the paths for the Berry-phase terms are $\text{C}_a=\{\bm{k}|k_z=-\pi,\ k_y=k_x+2\pi,\ -2\pi<k_x<0\}$
$\text{C}_b=\{\bm{k}|k_z=-\pi,\ k_y=-k_x+2\pi,\ 0<k_x<2\pi\}$, depicted in Fig.~\ref{fig:integral_region_9}(a). The term ${\tilde{F}}_{ij}\equiv
{\tilde{F}}_{ij}^{+}+ {\tilde{F}}_{ij}^{-}$ represents the total Berry curvature for the 
occupied states in SG $\bm{7}$.

Therefore, to summarize up to this point we have
\begin{align}
&\int_{\mathcal{A}}-\int_{\mathcal{C}}-
2\gamma_{EDE'}^{-}\\
&=-\frac{1}{2} \left( \int_{-\pi}^{\pi} dk_z \int_{-\pi}^{\pi} dk_x {F}_{zx} (k_x, \pi, k_z) 
\right.\nonumber\\
&+\left. \int_{-\pi}^{\pi} dk_z \int_{-\pi}^{\pi} dk_x {F}_{zx} (k_x, -\pi, k_z) \right) \nonumber \\
& + \sum_{i\in\text{occ.}} 
(\chi_i(\pi, \pi, -\pi) + \chi_i(-\pi, \pi, -\pi))
\nonumber\\
&+\int_{0}^{2\pi} dk_y \int_{k_y-2\pi}^{-k_y+2\pi} dk_x {F}_{xy} (k_x, k_y, -\pi) \nonumber \\
&  +\left(\oint_{\text{C}_a}+\oint_{\text{C}_b}\right)d\bm{k}\cdot \bm{A} (\bm{k}).
\label{eq:ACgamma}
\end{align}
In order to proceed, we have to clarify properties of the function $\chi_{i}(\bm{k})$. 
Owing to the periodicity of BZ,
we can write
\begin{align}
\chi_{i}(\bm{k} + \bm{b}_1^{Cc}) &= \chi_{i}(\bm{k}) + 2\pi N_i , \\
\chi_{i}(\bm{k} + \bm{b}_2^{Cc}) &= \chi_{i}(\bm{k}) + 2\pi N_i' ,
\end{align}
where $N_i$ and $N_i'$ are integers.
By substituting $\bm{k} \rightarrow \bm{k} + \bm{b}_1^{Cc}$ in Eq.~(\ref{eq:chi_rel_kz}),
we get $N_i=N_i'$. Then
we obtain an expression of $\chi_{i} (\bm{k})$ as
\begin{equation}
\chi_{i}(\bm{k}) = N_i k_y + \theta_i (k_x, k_y,k_z) ,
\end{equation}
where $N_i$ is an integer and $\theta_i (k_x, k_y,k_z)$ is a periodic function of $k_x$ and $k_y$ in the BZ of SG $\bm{{\it 9}}$:
$\theta_{i}(\bm{k} + \bm{b}_j^{Cc}) = \theta_{i}(\bm{k})$ ($j=1,2$).
Next, in Eq.~(\ref{eq:chi_rel}) we can write 
\begin{equation}
\chi_{i}(\bm{k}) + \chi_{i}(g_y \bm{k}) = - k_z+2\pi Q_i,
\label{eq:chi_rel_kz}
\end{equation}
where $Q_i$ is an integer.
Therefore, by summing over all the occupied states, we get
\begin{align}
&\sum_{i\in\text{occ.}} \left( \chi_i(\pi, \pi, -\pi) + \chi_i(-\pi, \pi, -\pi) \right) \nonumber\\
&= \sum_{i\in\text{occ.}} \left(2\pi N_i +
\theta_i(\pi, \pi, -\pi) + \theta_i(-\pi, \pi, -\pi) \right)\nonumber \\
&= \sum_{i\in\text{occ.}} \left(2\pi N_i +
\theta_i(\pi, \pi, -\pi) + \theta_i(\pi, -\pi, -\pi) \right)\nonumber\\
&= \sum_{i\in\text{occ.}} \left(2\pi N_i +2\pi Q_i+\pi
\right).
\label{eq:sumchi}
\end{align}
To further evaluate this quantity, we put $k_y=0$ in  Eq.~(\ref{eq:chi_rel_kz}), which 
leads to $2\chi_{i}(k_x,0,k_z) = - k_z+2\pi Q_i$. On the other hand, on the glide-invariant $k_y=0$ plane, 
the states are classified into the $g_{\pm}$ sectors with $G_y=\pm e^{-ik_z/2}$, meaning that
\begin{align*}
&\chi_{i}(k_x,0,k_z) \equiv - k_z/2 \pmod{2\pi} \ \rightarrow \ Q_i=\text{even}\ :\ g_+ \text{sector} \\
&\chi_{i}(k_x,0,k_z) \equiv \pi- k_z/2 \pmod{2\pi} \ \rightarrow \ Q_i=\text{odd}\ :\ g_- \text{sector}
\end{align*}
Since these two sectors switch when $k_z$ increases by $2\pi$, meaning that the numbers of
occupied states in the $g_+$ sector and that in the $g_-$ sector are the same. 
Therefore, we finally conclude from Eq.~(\ref{eq:sumchi}) that
\begin{align}
&\sum_{i\in\text{occ.}} \left( \chi_i(\pi, \pi, -\pi) + \chi_i(-\pi, \pi, -\pi) \right)
\nonumber\\
&\hspace{1.5cm} \equiv \sum_{i\in\text{occ.}} 2\pi N_i \pmod{4\pi}.
\label{eq:chi_prop}
\end{align}
Second, we investigate the Berry phase terms in Eq.~(\ref{eq:ACgamma}).
Due to the periodicity of the BZ, we have
\begin{align}
&  \oint_{\text{C}_b}d\bm{k}\cdot \bm{A} (\bm{k}) \nonumber\\
&= \sum_{i\in\text{occ.}} 2\pi N_i + \ \oint_{\text{C}_a}d\bm{k}\cdot \bm{A} (\bm{k}),
\label{eq:FA9-2}
\end{align}
Thus in the sum of Eqs.~(\ref{eq:chi_prop}) and (\ref{eq:FA9-2}), the $N_i$ terms becomes $4\pi$ times an integer.

On the other hand, the terms on the glide-invariant plane $k_y = 0$ in $\bm{{\it 9}}$ are given by
\begin{align}
&\int_{\mathcal{B}}=\int_{-\pi}^{\pi} dk_z \int_{-\pi}^{\pi} dk_x {\tilde{F}}_{zx}^- (k_x, 0, k_z)
\nonumber\\
& = \int_{-\pi}^{\pi} dk_z \int_{-2\pi}^{2\pi} dk_x {F}_{zx}^- (k_x, 0, k_z) ,
\label{eq:FB9-1}
\end{align}
and
\begin{equation}
\gamma_{A'BA}^+=\oint_{-\pi}^{\pi} dk_x {\tilde{A}}_x^+ (k_x, 0, -\pi) = \oint_{-2\pi}^{2\pi} dk_x {A}_x^+ (k_x, 0, -\pi) .
\label{eq:gamma-a-9}
\end{equation}

Therefore, by combining these results, the glide-$Z_2$ invariant for $\bm{{\it 9}}$ recast as follows:
\begin{align}
\nu
&= \frac{1}{2\pi} \left[ 2  \oint_{\text{C}_a}d\bm{k}\cdot\bm{A} (\bm{k})
\right.\nonumber\\
&-\left. 2 \oint_{-2\pi}^{2\pi} dk_x {A}_x^+ (k_x, 0, -\pi) \right. \nonumber \\
&  + \int_{0}^{2\pi} dk_y \int_{k_y-2\pi}^{-k_y+2\pi} dk_x {F}_{xy} (k_x, k_y, -\pi) 
\nonumber\\
&+ \int_{-\pi}^{\pi} dk_z \int_{-2\pi}^{2\pi} dk_x {F}_{zx}^- (k_x, 0, k_z)  \nonumber \\
&  - \frac{1}{2} \left( \int_{-\pi}^{\pi} dk_z \int_{-\pi}^{\pi} dk_x {F}_{zx} (k_x, \pi, k_z) \right.\nonumber\\
&+\left.\left. \int_{-\pi}^{\pi} dk_z \int_{-\pi}^{\pi} dk_x {F}_{zx} (k_x, -\pi, k_z) \right) \right] .
\label{eq:z2glide-9-1}
\end{align}
Because we are assuming an insulating system, the vector field of the Berry curvature is divergence free.
Therefore, by noticing that the summation of the last two terms in Eq.~(\ref{eq:z2glide-9-1}) is equal to the Chern number on the glide-invariant plane $k_y = 0$ 
(see Fig.~\ref{fig:integral_region_9}(b)), we have an alternative expression for $\nu$,
\begin{align}
\nu &= \frac{1}{2\pi} \left[ 2  \oint_{\text{C}_a}d\bm{k}\cdot \bm{A} (\bm{k})
\right.\nonumber\\
&\left.- 2\oint_{-2\pi}^{2\pi} dk_x {A}_x^+ (k_x, 0, -\pi) \right. \nonumber \\
&  \left. + \int_{0}^{2\pi} dk_y \int_{k_y-2\pi}^{-k_y+2\pi} dk_x {F}_{xy} (k_x, k_y, -\pi) 
\right.\nonumber\\
&\left. + \int_{-\pi}^{\pi} dk_z \int_{-2\pi}^{2\pi} dk_x  \frac{1}{2}({F}_{zx} (k_x, 0, k_z) - {F}_{zx}^+ (k_x, 0, k_z) )\right] ,
\label{eq:nu9}
\end{align}
which is the same as Eq.~(\ref{eq:z2glide-9-2}).
We have used the relation ${F}_{ij} (\bm{k}) = {F}_{ij}^+ (\bm{k}) + {F}_{ij}^- (\bm{k})$.

So far we studied spinless cases. Meanwhile, Eq.~(\ref{eq:nu9}) applies also to spinful cases as well. 
Instead of Eq.~(\ref{eq:Gu}), in spinful cases the glide operator satisties $G_y(g_y\bm{k})G_y(\bm{k})=-e^{-ik_z}$ and
$-e^{-ik_z} = e^{i\chi(\bm{k}) + i\chi(g_y \bm{k})}$ instead of Eq.~(\ref{eq:chi_rel}).
Therefore, instead of Eq.~(\ref{eq:evec_gpm}) the wavefunctions in each glide sector of $\bm{{\it 9}}$ is given by
\begin{widetext}
\begin{align}
\ket{\tilde{u}^\pm (k_x, \pi, k_z)} &= \frac{1}{2} \left[ 
\begin{pmatrix}
1 \\ i e^{ik_x/2} 
\end{pmatrix}
\ket{u (k_x, \pi, k_z)} \mp i
\begin{pmatrix}
1 \\ -i e^{ik_x/2}
\end{pmatrix} 
e^{i \chi(k_x, \pi, k_z) + ik_z/2} \ket{u (k_x, -\pi, k_z)} \right] \nonumber \\
& \quad \times \exp \left[ -i \left( \chi(\pi, \pi, k_z) + \frac{k_z}{2} \right) \frac{k_x}{2\pi} \right] \times
\begin{cases}
e^{i k_x/4} & (g_+ \ \mathrm{sector}) \\ e^{-i k_x/4} & (g_- \ \mathrm{sector})
\end{cases}
.
\end{align}
From this equation, the expression $\chi_i(\pi, \pi, -\pi) + \chi_i(-\pi, \pi, -\pi) 
$ in this Appendix acquires an additional $-\pi$ term. Nonetheless, Eq.~(\ref{eq:chi_rel_kz})
is replaced by 
\begin{equation}
\chi_{i}(\bm{k}) + \chi_{i}(g_y \bm{k}) =\pi - k_z+2\pi Q_i,
\end{equation}
which also has an additional $\pi$. 
As a result, Eq.~(\ref{eq:sumchi}) becomes
 \begin{align}
&\sum_{i\in\text{occ.}} \left( \chi_i(\pi, \pi, -\pi) + \chi_i(-\pi, \pi, -\pi) -\pi \right) \nonumber\\
&= \sum_{i\in\text{occ.}} \left(2\pi N_i +
\theta_i(\pi, \pi, -\pi) + \theta_i(-\pi, \pi, -\pi) -\pi \right)\nonumber \\
&= \sum_{i\in\text{occ.}} \left(2\pi N_i +
\theta_i(\pi, \pi, -\pi) + \theta_i(\pi, -\pi, -\pi)-\pi  \right)\nonumber\\
&= \sum_{i\in\text{occ.}} \left(2\pi N_i +2\pi Q_i+\pi
\right),
\end{align}
\end{widetext}
and its result is the same as the spinless cases. Thus, the final expression of the glide-$Z_2$ invariant in spinful cases is the same as that in spinless cases in  Eq.~(\ref{eq:nu9}).

\section{Layer construction for space groups $\bm{{\it 9}}$ and $\bm{{\it 15}}$}
\label{sec:app-LC}
In this appendix, we argue a layer construction (LC) in class A for $\bm{{\it 9}}$ and $\bm{{\it 15}}$.
The formalism for the LC in class A is the same as that in Ref.~\onlinecite{Kim2019prb100}.

\subsection{Setup}

Let us start with a layer $(mnl;d)$ which is a set of planes described by the Miller indices $(mnl)$ and displaced from the origin by $d$ $(0 \le d < 1)$,
\begin{align}
(mnl;d) &= \left\{ \bm{r} | \bm{r} \cdot (m \bm{b}_1 + n \bm{b}_2 + l \bm{b}_3) = 2\pi (d+q), \ q\in \mathbb{Z} \right\} \nonumber \\
&= \left\{ \bm{r} | \frac{mx}{a} + \frac{ny}{b} + \frac{lz}{c} = d+q, \ q\in \mathbb{Z} \right\} ,
\end{align}
where $\bm{b}_i$'s $(i = 1, 2, 3)$ are reciprocal vectors corresponding to lattice vectors $\bm{a}_1 = a(100), \bm{a}_2 = b(010), \bm{a}_3 = c(001)$ of the 
primitive lattice, which are $\bm{a}_i^{Pc}$'s in our main text. We take this choice for convenience.

\subsection{Topological invariants in glide-symmetric systems}

Here we summarize the results of the topological invariants 
associated with the layer construction comprising 2D Chern insulator layers.
The detailed explanations are in Ref.~\onlinecite{Kim2019prb100}, and we show its outline here.

\subsubsection{Chern invariant}
For a reciprocal plane $\mathcal{D}_i$ normal to one of the {\it primitive} lattice vectors $\bm{a}_i$, the Chern number is defined as
\begin{equation}
\delta_{\mathcal{C}_ i}  =\frac{1}{2\pi} \int_{\mathcal{D}_i} d^2 \bm{k} \ \bm{n}_i \cdot \bm{\Omega}_{\bm k} ,
\end{equation}
where $\bm{\Omega}_{\bm k}=(F_{yz},F_{zx},F_{xy})$ is the Berry curvature, $\bm{n}_i \equiv \bm{a}_i / | \bm{a}_i |$ is a unit vector along $\bm{a}_i$, and the integral is taken over the 2D BZ 
 $\mathcal{D}_i$ on the plane perpendicular to $\bm{n}_i$.

These Chern numbers are associated with the primitive lattice vectors $\bm{a}_i$, and we can calculate them by counting intersections of the LC $L = (mnl;d)$ with an axis parallel to the $\bm{a}_i$ within the unit cell.
Therefore, we have
\begin{equation}
\delta_{\mathcal{C}_ i}= \frac{1}{2\pi} \sum_{L \in E} \bm{\mathrm{g}}_L \cdot \bm{a}_i ,
\end{equation}
where $E$ is a set of LCs compatible with the space group considered, and $\bm{\mathrm{g}}_L = m\bm{b}_1 + n\bm{b}_2 + l\bm{b}_3$ for the layer $L \in E$.
We note that even though the Miller indices for the layer, the symmetry operations, and the other topological invariants are given by {\it conventional} lattice vectors, these Chern numbers are defined with respect to the {\it primitive} cell. Namely, in the SG $\bm{{\it 9}}$ and $\bm{{\it 15}}$ we take $\bm{a}_1^{Cc}=(\frac{1}{2},\frac{1}{2},0)$, 
$\bm{a}_2^{Cc}=(\bar{\frac{1}{2}},\frac{1}{2},0)$, $\bm{a}_3^{Cc}=(0,0,1)$.

\subsubsection{Glide invariant $\delta_{\mathrm g}$}
In the presence of glide symmetry, we can find a $Z_2$ invariant for glide-symmetric systems\cite{Fang2015prb91, Shiozaki2015prb91}.
We can express this invariant based on the LC in the similar way as in Ref.~\onlinecite{Song2018ncommun9}.
We first choose one of the glide planes, and we fix this choice throughout our theory.
Then, the glide invariant $\delta_{\mathrm{g}}$ is expressed as
\begin{align}
&\delta_{\mathrm{g}} = \sum_{L \in E} ( N^\mathrm{o}_{m, \bm{\mathrm{t}}_\parallel} (L) + N^\mathrm{s}_{m, \bm{\mathrm{t}}_\parallel} (L)) \pmod{2} \\
&N^\mathrm{o}_{m, \bm{\mathrm{t}}_\parallel} (L) =
\begin{cases}
1 & \mathrm{if} \ m \in L \\ 0 & \mathrm{otherwise}
\end{cases} , \\
&N^\mathrm{s}_{m, \bm{\mathrm{t}}_\parallel} (L) = \frac{1}{2\pi} | \bm{\mathrm{t}}_\parallel \cdot \bm{\mathrm{g}}_L | ,
\end{align}
where $m$ is the glide plane of our choice, $m \in L$ means the layer $L$ occupies the glide plane $m$, and $\bm{\mathrm{t}}_\parallel$ is a glide vector.

\subsubsection{Inversion invariant $\delta_{\mathrm i}$}

In the presence of inversion symmetry, we can define the inversion invariant $\delta_{\mathrm{i}}$.
We first choose one  out of the eight inversion centers in a unit cell, and we fix the choice throughout our theory.
When a layer occupies the inversion center, this layer cannot be trivialized as long as inversion symmetry is preserved.
Meanwhile, one can trivialize the configuration where two layers occupy the inversion center, rendering it into a trivial atomic insulator.
Therefore, the inversion invariant $\delta_{\mathrm{i}}$ is expressed as
\begin{align}
&\delta_{\mathrm{i}} (E) = \sum_{L \in E} N^\mathrm{o}_\mathrm{i} (L) \pmod{2}, \\
&N^\mathrm{o}_\mathrm{i} (L) =
\begin{cases}
1 & \mathrm{if} \ i \in L \\ 0 & \mathrm{otherwise}
\end{cases} ,
\end{align}
where $i$ is the inversion center of our choice out of the eight different inversion centers, $i \in L$ means that the layer $L$ occupies the inversion center $i$.

\subsubsection{$C_2$ rotation invariant $\delta_{\mathrm r}$}

In the presence of $C_2$ rotation symmetry, we obtain the $C_2$ rotation invariant $\delta_{\mathrm{r}}$ in a similar manner.
In fact, this invariant is always trivial as long as a layer is decorated by a 2D Chern insulator \cite{Kim2019prb100}.
Therefore, we have
\begin{equation}
\delta_{\mathrm{r}} (E) = 0.
\end{equation}

\subsubsection{$C_2$ screw invariant $\delta_{\mathrm s}$}

In the presence of $C_2$ screw symmetry, we can find the $C_2$ screw invariant $\delta_{\mathrm{s}}$ \cite{Kim2019prb100} as
\begin{align}
&\delta_{\mathrm{s}} (E) = \sum_{L \in E} N^\mathrm{s}_{s, \bm{\mathrm{t}}_\parallel} (L) \pmod{2} ,\\
&N^\mathrm{s}_{s, \bm{\mathrm{t}}_\parallel} (L) = \frac{1}{2\pi} | \bm{\mathrm{t}}_\parallel \cdot \bm{\mathrm{g}}_L | ,
\end{align}
where $\bm{\mathrm{t}}_\parallel$ is a screw vector.

\subsection{Invariants based on LCs}

\subsubsection{Space group $\bm{{\it 9}}$}

In $\bm{{\it 9}}$, we can find the Chern invariants $\delta_{\mathcal{C}_{i=1,2,3}}$, a glide invariant $\delta_{\mathrm{g}_y}$ defined by the glide operation $G_y = \{ M_y | \hat{\bm{z}} /2 \}$, and the other glide invariant $\delta_{\mathrm{g}_n}$ defined by the glide operation $G_n = \{ M_y | (\hat{\bm{x}}+\hat{\bm{y}}+\hat{\bm{z}}) /2 \}$.
The glide operation 
$G_n(=T_1G_y)$ is known as an {\it additional symmetry element} \cite{Hahn2002ITA}, where $T_1=\{E|(\hat{\bm{x}}+\hat{\bm{y}}) /2\}$, and the two glide operations lead to different glide invariants $\delta_{g_y}$ and $\delta_{g_n}$.

Let us investigate the invariants in $\bm{{\it 9}}$. The glide operations $G_y$ and $G_n$ transform the layer $L = (mnl;d)$ to
\begin{align}
G_y L &= \left( \bar{m} n \bar{l} \Big| -\frac{l}{2} - d \right), \\
G_n L &= \left( \bar{m} n \bar{l} \Big| -\frac{1}{2} (m-n+l) - d \right),
\end{align}
and
\begin{equation}
G_n G_y^{-1} L = T_1 L = \left( mnl; d +\frac{1}{2} (m+n) \right) .
\end{equation}
Note that $G_yL$ is not written as $(m\bar{n}l;d-\frac{l}{2})$ because the 2D Chern insulator has its own orientation corresponding to the direction of the chiral edge states.
Therefore, the Chern invariants for the LC $L(mnl;d)$ in $\bm{{\it 9}}$ are given as
\begin{align}
&\delta_{\mathcal{C}_1} = \delta_{\mathcal{C}_2}   , \\
&= \left\{
\begin{array}{ll}
n / 2 & (m = l = 0, n = \mathrm{even},d=0, 1/2) \\
n & (m = l = 0, n = \mathrm{odd},4d\in\mathbb{Z}  \  \\
 & \ \mathrm{or}\ m = l=0, n = \mathrm{even}, d\neq 0,1/2 \nonumber \\
 & \ \mathrm{or} \ (m,l)\neq (0,0), m+n = \mathrm{even}) \\
2n & (\mathrm{otherwise})
\end{array}
\right. , \\
&\delta_{\mathcal{C}_3} = 0 .
\end{align}
We emphasize that the Chern invariants $\delta_{\mathcal{C}_{i=1,2,3}}$ are associated with the {\it primitive} lattice vectors for $\bm{{\it 9}}$, $\bm{a}_i^{Cc}$ shown in  Table~\ref{table:prim_reci_vectors}.

The glide invariant $\delta_{\mathrm{g}_y}$ is given by
\begin{equation}
\delta_{\mathrm{g}_y} \equiv \left\{
\begin{array}{ll}
1 & (m = l =0, d = 0\ \text{or}\ \\
& \ \ m = l = 0, n=\text{odd}, d = 1/2) \\ l  & ((m,l)\neq (0,0), m+n = \mathrm{even}) \\ 0 & (\mathrm{otherwise})
\end{array}
\right.
\label{eq:LC9_bglide_invariant}
\end{equation}
modulo 2.
The glide plane is chosen to be $y = 0$, and not $y = 1/2$.
Similarly, the other glide invariant $\delta_{\mathrm{g}_n}$ is given by
\begin{equation}
\delta_{\mathrm{g}_n} \equiv \left\{
\begin{array}{ll}
1 & (m = l = 0, n \in 4\mathbb{Z}, d = 0 \ \mathrm{or} \\
 & \ m = l = 0, n = \mathrm{odd}, d = 1/4,3/4 \ \mathrm{or} \\
 & \ m = l = 0, n = 2\times \mathrm{odd}, d = 1/2) \\
m + l  & ((m,l)\neq (0,0), m+n = \mathrm{even}) \\ 
0 & (\mathrm{otherwise})
\end{array}
\right.
\end{equation}
modulo 2.
Here we choose the glide plane to be $y = 1/4$, and not $y = 3/4$.
One can directly show that $\delta_{g_n}\equiv\delta_{g_y}+\delta_{\mathcal{C}_1} \pmod{2}$. 

\subsubsection{Space group $\bm{{\it 15}}$}

In $\bm{{\it 15}}$, we can define six kinds of invariants, the Chern invariants $\delta_{\mathcal{C}_{i=1,2,3}}$, the two different types of glide invariant $\delta_{\mathrm{g}_y}$, $\delta_{\mathrm{g}_n}$, the inversion invariant $\delta_{\mathrm{i}}$, the $C_2$ rotation invariant $\delta_{\mathrm{r}}$, and the $C_2$ screw invariant $\delta_{\mathrm{s}}$
for $\{C_{2y}|(\hat{\bm{x}}+\hat{\bm{y}}+\hat{\bm{z}})/2\}$.
The Chern invariant $\delta_{\mathcal{C}_i}$ takes an integer value, while the other invariants $\delta_{\mathrm{g}_y}$, $\delta_{\mathrm{g}_n}$, $\delta_{\mathrm{i}}$, and $\delta_{\mathrm{s}}$ are defined modulo 2, and $\delta_{\mathrm{r}}$ turns out to be always zero.
The inversion operator transforms the layer $(mnl;d)$ into $(mnl;-d)$, and this leads to doubling to the number of layers to be eight, if the additional layers are not identical with the original ones.
In this system, the Chern invariants are obtained as follows:
\begin{align}
&\delta_{\mathcal{C}_1} =
\delta_{\mathcal{C}_2} \nonumber\\
&= \left\{
\begin{array}{ll}
n / 2 & (m = l = 0, n = \mathrm{even}, \\
 & d=0, 1/2) \\
n & (m =l =  0, n = \mathrm{odd}, \\ & d=0, 1/4, 1/2, 3/4 \\
n  & (m =l =  0, n = \mathrm{even},d\neq 0,1/2)\\
n & ((m,l)\neq (0,0),m + n = \mathrm{even}, d = 0, 1/2 ) \\
2n & ((m,l)\neq (0,0),m+n = \mathrm{even} , \ d \neq 0, 1/2 
\\& \  \mathrm{or} \ (m,l)\neq (0,0),m+n = \mathrm{odd}, 4d \in\mathbb{Z} 
 \\
& \  \mathrm{or} \ m =l = 0,n = \mathrm{odd}, d\neq 0, 1/4, 1/2, 3/4 ) \\
4n & (\mathrm{otherwise})
\end{array} \right. , \\
& \delta_{\mathcal{C}_3} = 0 .
\end{align}
The glide invariant $\delta_{\mathrm{g}_y}$ and the inversion invariant $\delta_{\mathrm{i}}$ are calculated, and they turn out to be completely equal;
\begin{align}
\delta_{\mathrm{g}_y} &= \delta_{\mathrm{i}} \equiv \left\{
\begin{array}{ll}
1 & (m =l = 0, d = 0\ \\
&\ \ \  \text{or}\ m = 0, n=\text{odd}, l =0, d = \frac{1}{2}) ) \\ 
l & ((m,l)\neq (0,0), m + n = \mathrm{even}, \\
& \ \ \ \ \ d = 0, 1/2) \\
0 & \mathrm{otherwise}
\end{array} \right.\\
\delta_{\mathrm{g}_y} &= \delta_{\mathrm{i}} 
\end{align}
modulo 2.
The other glide invariant $\delta_{\mathrm{g}_n}$ is given by
\begin{equation}
\delta_{\mathrm{g}_n} \equiv \left\{
\begin{array}{ll}
1 & (m = l = 0, n \in 4\mathbb{Z}, d = 0 \ \mathrm{or} \\
 & \ m = l = 0, n = \mathrm{odd}, d = 1/4,3/4 \ \mathrm{or} \\
 & \ m = l = 0, n/2 = \mathrm{odd}, d = \frac{1}{2}) \\
m + l  & (m+n = \mathrm{even}, d = 0, 1/2) \\ 
0 & (\mathrm{otherwise})
\end{array} \right.
\end{equation}
modulo 2.
The $C_2$ screw invariant $\delta_{\mathrm{s}}$ is given by
\begin{equation}
\delta_\mathrm{s} \equiv
\delta_{\mathcal{C}_1} = \delta_{\mathcal{C}_2}
\end{equation}
modulo 2.
They are related by the following equations:
\begin{align}
\delta_\mathrm{i} &\equiv \delta_{\mathrm{g}_y} \pmod{2} , \\
\delta_\mathrm{s} & \equiv \frac{1}{2} \delta_{\mathcal{C}_y} \pmod{2} , \\
\delta_\mathrm{i} + \delta_\mathrm{s} &\equiv \delta_{\mathrm{g}_n} \pmod{2} ,
\end{align}
in which we define $\delta_{\mathcal{C}_y} = \delta_{\mathcal{C}_1} + \delta_{\mathcal{C}_2} =2\delta_{\mathcal{C}_1}$.

Let us verify the relations between these invariants and the glide-$Z_2$ invariant $\nu$ and the Chern number $\mathcal{C}_y$.
We can calculate $\nu$ and $\mathcal{C}_y$ for general layers and compare them with the invariants from the LC.
We then get
\begin{align}
\delta_{\mathcal{C}_y} &= \mathcal{C}_y \in 2\mathbb{Z}, \\
\delta_{\mathrm{i}} \equiv \delta_{\mathrm{g}_y} &\equiv \nu \pmod{2}, \\
\delta_{\mathrm{s}} &\equiv \frac{1}{2} \mathcal{C}_y \pmod{2}, \\
\delta_{\mathrm{g}_n} &\equiv \nu + \frac{1}{2}\mathcal{C}_y \pmod{2} .
\end{align}
The topological invariants calculated from the LC hence totally agree with the known topological invariants,
the glide-$Z_2$ invariant $\nu$ and the Chern number $\mathcal{C}_y$.
We will show elementary layer constructions, showing minimal layer configurations with nontrivial topological invariants from these results shown in Table~\ref{table:eLCs}.

\subsection{Elementary layer constructions}
Following the argument in Ref.~\onlinecite{Song2018ncommun9}, we can find eLCs in $\bm{{\it 9}}$ and $\bm{{\it 15}}$ summarized in Table~\ref{table:eLCs}. 
All SGs listed here have two independent topological invariants 
as we have seen in this paper, Therefore, each SG has two eLCs.

\begin{table}[h!]
\caption{Summary of elementary layer construction, symmetry-based indicators, and topological invariants for $\bm{{\it 9}}$ and $\bm{{\it 15}}$.
$\delta_{\mathcal{C}_i}$ are Chern indices.  In addition, $\delta_{g_y}$, $ \delta_i$, $ \delta_{s}$ and $\delta_{g_n}$ are topological indices associated with
the glide $\{M_y|\hat{\bm{z}}/2\}$, inversion, $C_2$ screw $\{C_{2y}|(\hat{\bm{x}}+\hat{\bm{y}}+\hat{\bm{z}})/2\}$, and 
the glide $\{M_y|(\hat{\bm{x}}+\hat{\bm{y}}+\hat{\bm{z}})/2\}$, respectively.
In the notation of Ref.~\onlinecite{Song2018ncommun9}, 
these four symmetry operations are written as $\mathrm{g}^{010}_{00\frac{1}{2}}$, $i$, $2^{010}_{\frac{1}{2}\frac{1}{2}\frac{1}{2}}$, and $\mathrm{g}^{010}_{\frac{1}{2}\frac{1}{2}\frac{1}{2}}$.
}
$$
\begin{array}{c c c ccccc}
\hline \hline
\mathrm{SG} & \mathrm{eLC} & \mathbb{Z}_{2,2,2,4} & \multicolumn{5}{c}{\mathrm{Invariants}} \\ \hline
 							& (mnl;d) &  & \delta_{\mathcal{C}_i} & \delta_{g_y} & &  &\delta_{g_n}  \\ \hline
\bm{{\it 9}} \ \ (Cc) & 001; d_0 & \mathrm{N} \slash \mathrm{A} & 000 & 1 &   & &1 \\ 
 							& \bar{1}10; d_0 & & 110 & 0 &  &  &1 \\ \hline
 							& (mnl;d) & & \delta_{\mathcal{C}_i} & \delta_{g_y} & \delta_i &  \delta_{s}  &  \delta_{g_n}  \\ \hline
\bm{{\it 15}} \ (C2/c)  & 001; 0 & 0002 & 000 & 1 & 1 & 0 & 1\\
 								& \bar{1}10; 0 & 1100 & 110 & 0 & 0 &1 & 1 \\ 
\hline \hline
\end{array}
$$
\label{table:eLCs}
\end{table}

\section{Tight-binding models for the elementary layer constructions of $\bm{{\it 15}}$}
\label{app:layer15}

Let us consider the system with the SG $\bm{{\it 15}}$ with the glide symmetry and inversion symmetry given by 
\begin{equation}
G_y = \left\{ M_y  | \hat{\bm{z}}/2\right\} , \quad I = \{ I | 0 \} .
\end{equation}
As we presented in Appendix \ref{sec:app-LC}, there are two eLCs,
$(001;0)$ and $(\bar{1}10;0)$. We discuss these two eLCs separately. 

\subsection{$\nu = 1, \ \mathcal{C}_y = 0 \ (L = (001;0))$}
This eLC
 is given by putting a Chern insulator layer with $\mathcal{C} = 1$ on the $xy$ plane at $z = 0$ with the inversion symmetry, 
 and by generating other layers by symmetry
as shown in Fig~\ref{fig:lc15}(a). The model Hamiltonian for the layer on the $z=0$ plane is written as
\begin{align*}
&\left(\sum_{x,y \in \mathbb{Z}} + \sum_{x,y \in \mathbb{Z} + \frac{1}{2}} \right) \left[ \psi^\dagger (x+\frac{1}{2}, y+\frac{1}{2}) \frac{\sigma_z + i\sigma_x}{2} \psi (x, y) \right.\nonumber\\
&+ \psi^\dagger (x-\frac{1}{2}, y+\frac{1}{2}) \frac{\sigma_z + i\sigma_y}{2} \psi (x, y) + h.c.  \\
& + \left.m \psi^\dagger (x, y) \sigma_z \psi (x, y)\right] ,
\end{align*}
satisfying
\begin{equation*}
\hat{I} \psi^\dagger (x, y) \hat{I}^{-1} = \psi^\dagger (-x, -y) \sigma_z .
\end{equation*}
Making copies by the glide transformation
\begin{equation}
\hat{G}_y \psi^\dagger (x, y, z) \hat{G}_y^{-1} = \psi^\dagger (x, -y, z + \frac{1}{2}) .
\end{equation}
\begin{widetext}
The Hamiltonian reads
\begin{align*}
\hat{H} &= \left(\sum_{x,y,z \in \mathbb{Z}} + \sum_{x,y \in \mathbb{Z} + \frac{1}{2}, z \in \mathbb{Z}} \right) \left[ \psi^\dagger (x+\frac{1}{2}, y+\frac{1}{2}, z) \frac{\sigma_z + i\sigma_x}{2} \psi(x, y, z) + \psi^\dagger (x-\frac{1}{2}, y+\frac{1}{2}, z) \frac{\sigma_z + i\sigma_y}{2} \psi(x, y, z) \right. \\
&+  \psi^\dagger (x+\frac{1}{2}, -y-\frac{1}{2}, z+\frac{1}{2}) \frac{\sigma_z + i\sigma_x}{2} \psi(x, -y, z+\frac{1}{2})+ \psi^\dagger (x-\frac{1}{2}, -y-\frac{1}{2}, z+\frac{1}{2}) \frac{\sigma_z + i\sigma_y}{2} \psi (x, -y, z+\frac{1}{2}) + h.c. \\
&+m \psi^\dagger (x, y, z) \sigma_z \psi (x, y, z)+ \left.m \psi^\dagger (x, -y, z+\frac{1}{2}) \sigma_z \psi (x, -y, z+\frac{1}{2}) \right]\\
&= \left(\sum_{x,y,z \in \mathbb{Z}} + \sum_{x,y \in \mathbb{Z} + \frac{1}{2}, z \in \mathbb{Z}} \right) \left[ \psi^\dagger (x+\frac{1}{2}, y+\frac{1}{2}, z) \frac{\sigma_z + i\sigma_x}{2} \psi(x, y, z) + \psi^\dagger (x-\frac{1}{2}, y+\frac{1}{2}, z) \frac{\sigma_z + i\sigma_y}{2} \psi(x, y, z)  \right. \\
&+  \psi^\dagger (x+\frac{1}{2}, y-\frac{1}{2}, z+\frac{1}{2}) \frac{\sigma_z + i\sigma_x}{2} \psi(x, y, z+\frac{1}{2}) + \psi^\dagger (x-\frac{1}{2}, y-\frac{1}{2}, z+\frac{1}{2}) \frac{\sigma_z + i\sigma_y}{2} \psi(x, y, z+\frac{1}{2}) + h.c. \\
&+  m \psi^\dagger (x, y, z) \sigma_z \psi (x, y, z)+ \left. m \psi^\dagger (x, y, z+\frac{1}{2}) \sigma_z \psi (x, y, z+\frac{1}{2})\right] \end{align*}
Let us introduce $k$-space basis
\begin{equation}
\Psi^\dagger (k_x, k_y, k_z) \equiv \left( \sum_{x, y, z \in \mathbb{Z}} + \sum_{x, y \in \mathbb{Z} + \frac{1}{2}, z \in \mathbb{Z}} \right)  \left( \psi^\dagger (x, y, z), \psi^\dagger(x, y, z+\frac{1}{2}) \right) e^{i(k_x x + k_y y + k_z z)} .
\end{equation}
The glide and inversion are replaced as
\begin{align*}
\hat{G}_y \Psi^\dagger (k_x, k_y, k_z) \hat{G}_y^{-1} &= \left( \sum_{x, y, z \in \mathbb{Z}} + \sum_{x, y \in \mathbb{Z} + \frac{1}{2}, z \in \mathbb{Z}} \right) \left( \psi^\dagger (x, -y, z+\frac{1}{2}), \psi^\dagger (x, -y, z+1) \right) e^{i(k_x x + k_y y + k_z z)} \\
&= \Psi^\dagger (k_x, -k_y, k_z)
\begin{pmatrix} 0 & e^{-ik_z} \\ 1 & 0 \end{pmatrix} _\tau , \\
\hat{I} \Psi^\dagger (k_x, k_y, k_z) \hat{I}^{-1} &= \left( \sum_{x, y, z \in \mathbb{Z}} + \sum_{x, y \in \mathbb{Z} + \frac{1}{2}, z \in \mathbb{Z}} \right)  \left( \psi^\dagger (-x, -y, -z) , \psi^\dagger (-x, -y, -z-\frac{1}{2}) \right) e^{i(k_x x + k_y y + k_z z)} \sigma_z \\
&= \Psi^\dagger (-k_x, -k_y, -k_z) \sigma_z \begin{pmatrix} 1 & 0 \\ 0 & e^{-ik_z}  \end{pmatrix} _\tau .
\end{align*}
To summarize, the momentum space Hamiltonian and symmetry operator are
\begin{align*}
H(k_x, k_y, k_z) &= \left( m + \cos \frac{k_x + k_y}{2} + \cos \frac{-k_x + k_y}{2} \right)\sigma_z \tau_0 
+ \sin \frac{k_x + k_y}{2} \begin{pmatrix} \sigma_x & 0 \\ 0 & -\sigma_y \end{pmatrix} 
+ \sin \frac{k_x - k_y}{2} \begin{pmatrix} -\sigma_y & 0 \\ 0 & \sigma_x \end{pmatrix} \\
&= \left( m + 2 \cos \frac{k_x}{2} \cos \frac{k_y}{2} \right) \sigma_z \tau_0 + \sin \frac{k_x}{2} \cos \frac{k_y}{2} (\sigma_x - \sigma_y) \tau_0 + \cos \frac{k_x}{2} \sin \frac{k_y}{2} (\sigma_x + \sigma_y) \tau_z ,
\end{align*}
\begin{align*}
G_y (k_x, k_y, k_z) = \begin{pmatrix} 0 & e^{-ik_z} \\ 1 & 0 \end{pmatrix} _\tau , \quad
I(k_x, k_y, k_z) = \sigma_z \begin{pmatrix} 1 & 0 \\ 0 & e^{-ik_z} \end{pmatrix} _\tau .
\end{align*}
\end{widetext}
Let us compute the indicator for $-2 < m < 0$. The $C_2$ rotation around the axis $x=0$, $z=1/4$ is
given by 
\begin{equation}
C_2 (k_x, k_y, k_z) \equiv G_y(-k_x, -k_y, -k_z) I(k_x, k_y, k_z) = \sigma_z \tau_x
\end{equation}
At high-symmetry points, the Hamiltonians and symmetry operators within the occupied states are given by the following:
$$
\begin{array}{c|cccc}
P & H(P) & C_2(P) |_{\mathrm{occ}} & G_y(P) |_{\mathrm{occ}} & I(P) |_{\mathrm{occ}} \\ \hline
\Gamma & (m+2) \sigma_z & \sigma_z\tau_x & \tau_x & \sigma_z \\
Y & (m-2) \sigma_z & \sigma_z\tau_x & \tau_x & \sigma_z \\
V & m \sigma_z & \sigma_z\tau_x & \tau_x & \sigma_z
\end{array}
$$
Thus the irreps for various values of $m$ in insulating cases ($m\neq 0,\pm 2$) are summarized in Table \ref{table:LCirreps15}.
Therefore, we find that the glide-$Z_2$ invariant is $\nu = 1$ when $0<|m|<2$ and $\nu = 0$ when $|m|>2$, in agreement with the expected results.

\subsection{$\nu = 0, \ \mathcal{C}_y = 2 \ (L = (\bar{1}10;0))$ }

We here calculate the topological invariants for the LC $(\bar{1}10;0)$
as shown in Fig.~\ref{fig:lc15}(b).
This layer is accompanied by the layer $(110;0)$ by $C_{2z}$ symmetry.
We consider a Chern insulator with $\mathcal{C} = 1$ on a $[\bar{1}10]$ plane including the origin.
Its Hamiltonian and the inversion operation is written as
\begin{align}
&\left(\sum_{x,y,z \in \mathbb{Z}} + \sum_{x,y \in \mathbb{Z} + \frac{1}{2}, z \in \mathbb{Z}} \right) \left[ \psi^\dagger (x, y, z+1) \frac{\sigma_z + i \sigma_x}{2} \psi(x, y, z) 
\right.\nonumber\\
&+ \psi^\dagger (x + \frac{1}{2}, y + \frac{1}{2}, z) \frac{\sigma_z + i\sigma_y}{2} \psi(x, y, z) + h.c.
\nonumber\\
&+ \left. m \psi^\dagger (x, y, z) \sigma_z \psi (x, y, z) \right]  ,
\end{align}
\begin{equation}
\hat{I} \psi^\dagger (x, y, z) \hat{I}^{-1} = \psi^\dagger (-x, -y, -z) \sigma_z .
\end{equation}
We then make its copies by the glide transformation
\begin{equation}
\hat{G}_y \psi^\dagger (x, y, z) \hat{G}_y^{-1} = \psi(x, -y, z+\frac{1}{2}) .
\end{equation}
\begin{widetext}
Then the total Hamiltonian reads
\begin{align*}
\hat{H} &= \left(\sum_{x,y,z \in \mathbb{Z}} + \sum_{x,y \in \mathbb{Z} + \frac{1}{2}, z \in \mathbb{Z}} \right) \left[ \psi^\dagger (x, y, z+1) \frac{\sigma_z + i\sigma_x}{2} \psi(x, y, z) + \psi^\dagger (x+\frac{1}{2}, y+\frac{1}{2}, z) \frac{\sigma_z + i\sigma_y}{2} \psi(x, y, z)  \right. \\
&+ \psi^\dagger (x, -y, z+\frac{3}{2}) \frac{\sigma_z + i\sigma_x}{2} \psi(x, -y, z+\frac{1}{2}) + \psi^\dagger (x+\frac{1}{2}, -y-\frac{1}{2}, z+\frac{1}{2}) \frac{\sigma_z + i\sigma_y}{2} \psi (x, -y, z+\frac{1}{2}) + h.c. \\
&+ m \psi^\dagger (x, y, z) \sigma_z \psi (x, y, z) +\left. m \psi^\dagger (x, -y, z+\frac{1}{2}) \sigma_z \psi (x, -y, z+\frac{1}{2}) \right]\\
&= \left(\sum_{x,y,z \in \mathbb{Z}} + \sum_{x,y \in \mathbb{Z} + \frac{1}{2}, z \in \mathbb{Z}} \right) \left[ \psi^\dagger (x, y, z+1) \frac{\sigma_z + i\sigma_x}{2} \psi(x, y, z) + \psi^\dagger (x+\frac{1}{2}, y+\frac{1}{2}, z) \frac{\sigma_z + i\sigma_y}{2} \psi(x, y, z)  \right. \\
&+ \psi^\dagger (x, y, z+\frac{3}{2}) \frac{\sigma_z + i\sigma_x}{2} \psi(x, y, z+\frac{1}{2}) + \psi^\dagger (x+\frac{1}{2}, y-\frac{1}{2}, z+\frac{1}{2}) \frac{\sigma_z + i\sigma_y}{2} \psi (x, y, z+\frac{1}{2}) + h.c.  \\
&+ m \psi^\dagger (x, y, z) \sigma_z \psi (x, y, z) +\left.m \psi^\dagger (x, y, z+\frac{1}{2}) \sigma_z \psi (x, y, z+\frac{1}{2}) \right].
\end{align*}

\end{widetext}

Let us introduce the $k$-space basis
\clearpage
\begin{align}
&\Psi^\dagger (k_x, k_y, k_z) =\left(\sum_{x,y,z \in \mathbb{Z}} + \sum_{x,y \in \mathbb{Z} + \frac{1}{2}, z \in \mathbb{Z}} \right) \nonumber\\
&\ \ \ 
( \Psi^\dagger (x, y, z), \Psi^\dagger (x, y, z+\frac{1}{2})) e^{i (k_x x + k_y y + k_z z)} .
\end{align}
The momentum space Hamiltonian and symmetry operators are represented as
\begin{align*}
&H(\bm{k}) = 
\begin{pmatrix}
H^{(+)}(\bm{k}) & \\ & H^{(-)}(\bm{k})
\end{pmatrix} _{\tau} \end{align*}
\begin{align*}
&H^{(\pm)} (\bm{k}) = \left( m + \cos k_z + \cos \frac{k_x \pm k_y}{2} \right) \sigma_z \nonumber\\
&+ \sin k_z \sigma_x + \sin \frac{k_x \pm k_y}{2} \sigma_y ,
\end{align*}
and
\begin{align*}
&G_y (k_x, k_y, k_z) = \begin{pmatrix} 0 & e^{-ik_z} \\ 1 & 0 \end{pmatrix} _\tau , \\
&I(k_x, k_y, k_z) = \sigma_z \begin{pmatrix} 1 & 0 \\ 0 & e^{-ik_z} \end{pmatrix} _\tau .
\end{align*}
The $C_2$ operation is 
\begin{equation}
C_2 (k_x, k_y, k_z) \equiv G_y (-k_x, -k_y, -k_z) I (k_x, k_y, k_z) = \sigma_z \tau_x.
\end{equation}
Let us compute the indicator for $-2 < m < 0$.
At the high-symmetry points, the Hamiltonian and the symmetry operators within the occupied states are given by the following:
$$
\begin{array}{c|ccc}
P & H(P) & G_y |_{\mathrm{occ}} & I |_{\mathrm{occ}} \\ \hline
\Gamma & (m+2) \sigma_z & \tau_x & \sigma_z \\
Y & m \sigma_z & \tau_x & \sigma_z \\
V (V^\prime) & (m+1) \sigma_z \mp \sigma_z\tau_z & & \sigma_z \\
M & (m-2) \sigma_z & -i \tau_y & \sigma_z\tau_z \\
L (L^\prime) & (m-1) \sigma_z \mp \sigma_z\tau_z & & \sigma_z\tau_z \\
\end{array}
$$
Thus the irreps for various values of $m$ in insulating cases ($m\neq 0,\pm 2$) are summarized in Table \ref{table:LCirreps15}.
It confirms that $\mathcal{C}_y/2 = 1 \pmod{2}$ when  $|m|<2$ and $\mathcal{C}_y/2 = 0 \pmod{2}$ when $|m|>2$, whereas $\nu$ is trivial on both sides, in agreement with the expected results.

\section{Table for the irreps of the SG $\bm{{\it 15}}$}
\label{sec:irrep15}
Here we summarize  the irreps of the SG $\bm{{\it 15}}$
used throughout the present paper in Table \ref{table:total_irreps15}.

\begin{widetext}
\begin{table}[htb]
\centering
\caption{Summary of irreducible representations, correspondences to characters denoted in the main text and the numbers of states in the irreps for $\bm{{\it 15}}$ where $a, b, m, x, y$, and $z$ are  integers. 
The number of states in this table represents the compatibility relations.
The positions of the high-symmetry points and lines are 
$\Gamma=(0,0,0)$, $Y=(0,2\pi,0)$, $V=(\pi,\pi,0)$, $A=(0,0,\pi)$, $M=(0,2\pi,\pi)$, $L=(\pi,\pi,\pi)$, $\Lambda=(0,v,0)$, and $B=(u,0,w)$.}
$$
\begin{array}{c c c c c c}

\hline\hline
\mathrm{Seitz} & \{ 1 | t_1, t_2, t_3 \} & \{ 2_{010} | 0, 0, 1/2 \} & \{ \bar{1} | 0, 0, 0 \} & \{ m_{010} | 0, 0, 1/2 \} & 
\begin{matrix}
\mathrm{number} \\ \mathrm{of} \\ \mathrm{states}
\end{matrix} \\ \hline
\begin{matrix}
\mathrm{Matrix} \\ \mathrm{presentation}
\end{matrix}  &
\begin{pmatrix}
1 & 0 & 0 & t_1 \\ 0 & 1 & 0 & t_2 \\ 0 & 0 & 1 & t_3
\end{pmatrix} & 
\begin{pmatrix}
-1 & 0 & 0 & 0 \\ 0 & 1 & 0 & 0 \\ 0 & 0 & -1 & 1/2
\end{pmatrix} &
\begin{pmatrix}
-1 & 0 & 0 & 0 \\ 0 & -1 & 0 & 0 \\ 0 & 0 & -1 & 0
\end{pmatrix} &
\begin{pmatrix}
1 & 0 & 0 & 0 \\ 0 & -1 & 0 & 0 \\ 0 & 0 & 1 & 1/2
\end{pmatrix} & \\ \hline
\Gamma^+_1 & 1 & 1 & 1 & 1 & a \\
\Gamma^-_1 & 1 & 1 & -1 & -1 & b+m\\
\Gamma^+_2 & 1 & -1 & 1 & -1 & a-m \\
\Gamma^-_2 & 1 & -1 & -1 & 1 & b \\[4pt]
\mathrm{Y}^+_1 & e^{i 2\pi t_2} & 1 & 1 & 1 & a+x \\
\mathrm{Y}^-_1 & e^{i 2\pi t_2} & 1 & -1 & -1 & b+m-x \\
\mathrm{Y}^+_2 & e^{i 2\pi t_2} & -1 & 1 & -1 & a-m+x \\
\mathrm{Y}^-_2 & e^{i 2\pi t_2} & -1 & -1 & 1 & b-x \\[4pt]
\mathrm{V}^+_1 & e^{i\pi (t_1 + t_2)} &  & 1 &  & a+b+y \\
\mathrm{V}^-_1 & e^{i\pi (t_1 + t_2)} &  & -1 &  & a+b-y \\[4pt]
\mathrm{A}_1 & e^{i\pi t_3} \sigma_0 & \sigma_x & \sigma_z & -i \sigma_y & 2(a+b) \\[4pt]
\mathrm{M}_1 & e^{i\pi (2t_2+t_3)} \sigma_0 & \sigma_x & \sigma_z & -i \sigma_y & 2(a+b) \\[4pt]
\mathrm{L}^+_1 & e^{i\pi (t_1+t_2+t_3)} &  & 1 &  & a+b+z \\
\mathrm{L}^-_1 & e^{i\pi (t_1+t_2+t_3)} &  & -1 &  & a+b-z \\[4pt]
\Lambda_1 & e^{i2\pi t_2 v} & 1 & & & a+b+m \\
\Lambda_1 & e^{i2\pi t_2 v} & -1 & & & a+b-m \\[4pt]
\mathrm{B}_1 & e^{i2\pi (t_1 u+t_3 w)} & & & e^{i\pi w} & a+b \\
\mathrm{B}_2 & e^{i2\pi (t_1 u+t_3 w)} & & & e^{i\pi (1+w)} & a+b \\[4pt]
\hline\hline
\end{array}
$$
\label{table:total_irreps15}
\end{table}

\end{widetext}

\clearpage



%

\end{document}